\documentclass[aps,twocolumn,showpacs,pra,superscriptaddress,floatfix,longbibliography]{revtex4-2}

\usepackage{amsmath,amsthm,amsfonts,amssymb,times,bbm,graphicx,color,epsfig,bm,soul}
\usepackage{hyperref}
\hypersetup{
   colorlinks=true,        
   linkcolor=blue,         
   citecolor=blue,         
   filecolor=blue,         
   urlcolor=blue           
}
\usepackage[normalem]{ulem}

\newcommand{\bra}[1]{\mbox{$\langle #1 |$}}
\newcommand{\ket}[1]{\mbox{$| #1 \rangle$}}

\begin{document}

\title{How creating one additional well can generate Bose-Einstein condensation}

\author{Mih\'aly M\'at\'e}
\affiliation{Wigner Research Centre for Physics, Hungarian Academy of Sciences, Konkoly-Thege Mikl\'os \'ut 29-33, 1121 Budapest, Hungary}
\affiliation{Department of Physics of Complex Systems, E\"otv\"os Lor\'and University, Pf. 32, H-1518 Budapest, Hungary}

\author{\"Ors Legeza}
\affiliation{Wigner Research Centre for Physics, Hungarian Academy of Sciences, Konkoly-Thege Mikl\'os \'ut 29-33, 1121 Budapest, Hungary}

\author{Rolf Schilling}
\affiliation{Institut f\"ur Physik, Johannes Gutenberg-Universit\"at, D-55099 Mainz, Germany}

\author{Mason Yousif}
\affiliation{Clarendon Laboratory, University of Oxford, Parks Road, Oxford OX1 3PU, United Kingdom}

\author{Christian Schilling}
\email{c.schilling@physik.uni-muenchen.de}
\affiliation{Department of Physics, Arnold Sommerfeld Center for Theoretical Physics,
Ludwig-Maximilians-Universit\"at M\"unchen, Theresienstrasse 37, 80333 M\"unchen, Germany}
\affiliation{Clarendon Laboratory, University of Oxford, Parks Road, Oxford OX1 3PU, United Kingdom}

\date{\today}

\begin{abstract}

\noindent\textbf{\large{Abstract}}\\
The realization of Bose-Einstein condensation in ultracold trapped gases has led to a revival of interest in that fascinating quantum phenomenon. This experimental achievement necessitated both extremely low temperatures and sufficiently weak interactions.
Particularly in reduced spatial dimensionality even an infinitesimal interaction immediately leads to a departure to quasi-condensation.
We propose a system of strongly interacting bosons which overcomes those obstacles by exhibiting a number of intriguing related features: (i) The tuning of just a single control parameter drives a transition from quasi-condensation to complete condensation, (ii) the destructive influence of strong interactions is compensated by the respective increased mobility, (iii) topology plays a crucial role since a crossover from one- to `infinite'-dimensionality is simulated, (iv) a ground state gap opens which makes the condensation robust to thermal noise. Remarkably, all these features can be derived by analytical and exact numerical means despite the non-perturbative character of the system.
\end{abstract}


\maketitle

\vspace{0.3cm}
\noindent\textbf{\large{Introduction}}\\
Bose-Einstein condensation(BEC) is one of the most striking quantum phenomena in nature \cite{Griffin1995,Pethick2002,Leggett2006,Pitaevskii2016}. While its theoretical prediction
dates back almost one hundred years ago it has more recently seen a revival of interest due to its realization in trapped
gases\cite{Anderson1995,Bradley1995,Davis1995}.
The accurate study of BEC by theoretical and computational approaches
particularly for systems with strong quantum correlations is rather challenging. This has been the reason why most studies of BEC so far were concerned with weakly interacting bosons(corresponding to the  experimental situation for ultracold gases) or even ideal Bose gases, eventually allowing for feasible mean-field approaches. Prime examples  are the Bogoliubov theory\cite{Bogoliubov1947} for uniform systems, Gross-Pitaevskii theory\cite{Gross1961,Gross1963,Pitaevskii1961} for general inhomogeneous systems, and perturbation theoretical approaches \cite{Lee1957a,Lee1957b,Brueckner1957a,Brueckner1957b,Beliaev1958,Hugenholtz1959,Lieb1963c}.
Although these widely used approaches have led to a deeper understanding of BEC, their range of validity is limited. To go beyond that limitation, various methods were developed\cite{Dalfovo1999,Andersen2004,Cazalilla2011,Lode2020}.

Since the experimental realization of BEC, the respective field of ultracold gases has become
one of the most exciting fields of research with a fruitful interplay between theory and experiment. It allowed for the experimental verification of numerous other theoretical predictions as well, stimulated further theoretical investigations of trapped particles \cite{Dalfovo1999}  and even revealed phenomena not observed before such as the crossover from BEC-superfluidity to BCS-superconductivity \cite{Greiner2003,Bartenstein2004,Zwierlein2004,Bourdel2004}. One of the most promising recent avenues has been the study of effectively one-dimensional quantum systems
\cite{Greiner2001,Dettmer2001,Goerlitz2001,Orzel2001,Paredes2004,Kinoshita2004,Stoeferle2004}.
Their most striking difference to three-dimensional systems is probably the absence of BEC: Already an infinitesimally weak interaction between the $N$ bosons leads to a ``sublinear'' behaviour of the number of condensed bosons, $N_0(N) \sim N^{\alpha}$\cite{Lenard1964,Popov1972}, even at zero temperature, for homogeneous gases as well for gases in a harmonic trap and regardless of the form of the interaction \cite{Popov1972,Widom1973,Schwartz1977,Girardeau2001,Forrester2003b,Gangardt2004,RigolHCB1d2004,RigolHCB1dPRL}. A prominent system giving rise to this  phase called ``quasi-condensation'' \cite{Popov1972}
is the Lieb-Liniger model\cite{Lieb1963a,Lieb1963b}, a ring system with $N$ spinless bosons interacting via a $\delta$-potential. Tuning the coupling constant to infinity leads to impenetrable bosons (Tonk-Girardeau gas)\cite{Girardeau1960} with the proven scaling $N_0(N) \sim \sqrt{N}$\cite{Lenard1964}.

Thermodynamic phase transitions(at finite temperatures)  in $D=3$ dimensions have been studied for more than a century. However, the study of quantum phase transitions(at zero temperature) \cite{Sachdev1999}, and particularly of the entanglement close to that transition \cite{Osterloh2002,Osborne2002} have attracted much attention in recent years, only. The latter studies were performed mostly for low-dimensional lattice models. They have revealed a striking similarity between the behavior of the order parameter  and of quantum informational quantities, like entanglement entropy.
 As discussed above, at zero-temperature  an interacting  Bose gas exhibits two qualitatively different phases, a quasi-condensate in $D=1$ and a true BE-condensate in $D \geq 3 $.  Therefore, it is of interest to search for a model which exhibits a  transition(or a crossover)  between these two phases, and in particular allows to check whether this special transition has common properties with  general quantum phase transitions.

BEC was  explored in cylindrical or torroidal trap geometries, both experimentally \cite{Greiner2001,Dettmer2001,Goerlitz2001,Orzel2001,Paredes2004,Kinoshita2004,Stoeferle2004} and theoretically \cite{Das2002,Salasnich2004, Salasnich2005}. But, changing the radial dimension of the confinement, neither the transition  from the sublinear $N$-dependence of $N_0(N) $ of the quasi-condensate to the linear dependence of the true condensate, nor its entanglement properties have been investigated. The only systematic study of such a transition was performed for a one-dimensional Bose gas in a harmonic trap \cite{Petrov2000}. However, that transition occurs only at temperatures  $T > 0$.

It is the challenge of the present work to propose and investigate a  lattice model for strongly interacting bosons
which allows one to drive such a transition by  changing just a single parameter, $s/t$, which is the ratio of the model's two hopping rates $s$ and $t$, as explained below. One of our major results is to establish by this model a  mechanism which can generate `infinite' range hopping by increasing  $s/t$. This is important since enhancing the  boson's mobility  allows to overcompensate the destructive effects of the repulsive interactions, leading finally to maximal possible condensation, despite infinitely strong repulsion. A further important feature of our model is  the generation  of an excitation gap in the $N$-particle spectrum for $s/t > 0$. This makes  BEC even robust to thermal noise and quantum fluctuations and thus may allow experimentalists to overcome the typical obstacles faced while realizing BEC.
The other important result concerns the application of tools from quantum information theory.  We show that the mutual information possesses the qualitatively similar dependence on $s/t$ as the number $N_0(N)$ of condensed bosons.  This supports the connection between the behavior of an order parameter and of entanglement at a  quantum phase transition even for the transition(or crossover) from a quasi-condensate to a true one.


All these key findings will be derived by analytical or exact numerical means despite the nonperturbative character of our system.

\vspace{0.3cm}
\noindent\textbf{\large{Results}}

\noindent\textbf{Model Hamiltonian.} To motivate our model, let us first recall that the possible presence of BEC depends in general not only on
the spatial dimensionality and temperature but  also on the ratio between  kinetic and interaction energy. In case of systems which are inhomogeneous, e.g., due to the presence of an external field or disorder, the occurrence of BEC will also depend on these quantities.
Concerning the  ratio between  kinetic and interaction energy, lattice systems have the great advantage that the kinetic energy can be manipulated by varying the hopping range between the lattice sites.
The most prominent lattice model for bosons is the widely studied
Bose-Hubbard model \cite{Fisher1989}
\begin{equation}\label{eq:BH-model}
\hat{H}= -\sum_{i,j} t_{ij} b_{i}^{\dagger} b_{j} +   U \sum_{i=1} \hat{n}_{i}(\hat{n}_{i}-1),
\end{equation}
where $b_{i}^{\dagger},  b_{i}$ creates/annihilates a spinless boson at site $i$, $\hat{n}_{i}\equiv b_{i}^{\dagger} b_{i}$ and $t_{ij}$ is the hopping rate between sites $i$ and $j$.  It was shown that the Bose-Hubbard model can be experimentally  realized by ultracold bosonic atoms in an  optical lattice \cite{Jaksch1998}.

The conflict between interaction and mobility is maximized in the limit of strong interactions $U \rightarrow \infty$ in which the bosons become hard-core\cite{Matsubara1956,Matsuda1957}. By employing respective hard-core boson(HCB) creation($h_{i}^{\dagger}$) and annihilation operators($ h_{i}$)  \eqref{eq:BH-model} takes the compact form $\hat{H}_{hc}= -\sum_{i,j} t_{ij} h_{i}^{\dagger} h_{j}$. Particularly the case of HCBs makes clear the important role of the hopping range, since for infinite-range hopping(a kind of mean field limit \cite{Fisher1989}) HCBs exhibit BEC even at finite temperatures, despite their infinitely strong repulsion \cite{Toth1990,Penrose1991,Kirson2000}.

Moreover, the effect of the interaction on BEC is distinctively destructive in one-dimensional systems. At zero temperature even an infinitesimally weak interaction already leads to a departure from BEC to the phase of quasi-condensation. This raises a fundamental question which our work shall answer in an affirmative and constructive way: After having confined a 3D Bose gas to one dimension, is it possible to tweak in an experimentally feasibly way this one-dimensional system with the effect of enhancing the mobility of the interacting bosons to reintroduce BEC? From a general point of view, one is immediately tempted to negate this question. The hopping amplitudes $t_{ij}$ namely resemble the overlap of Wannier orbitals at sites $i,j$ which in turn decays exponentially as function of the spatial separation $|i-j|$. Screening effects reduce the hopping even further and eventually motivate the common restriction of $t_{ij}$ in the Bose-Hubbard model to just nearest neighbors. The potential physical significance of long-range hopping has motivated experimentalists in recent years to realize at least effectively hopping terms beyond nearest neighbors. Despite a remarkable effort, the regime of infinite-range hopping has been out of reach but only the typical decay of $t_{ij}$ could be slowed down to an algebraic dipolar- and van der Walls-type one \cite{Guenter2013,Schempp2015}. It will be  one of our key achievements to propose a model which eventually would allow one to enhance the mobility even to infinite-range.

\begin{figure}
 \includegraphics[width=7cm]{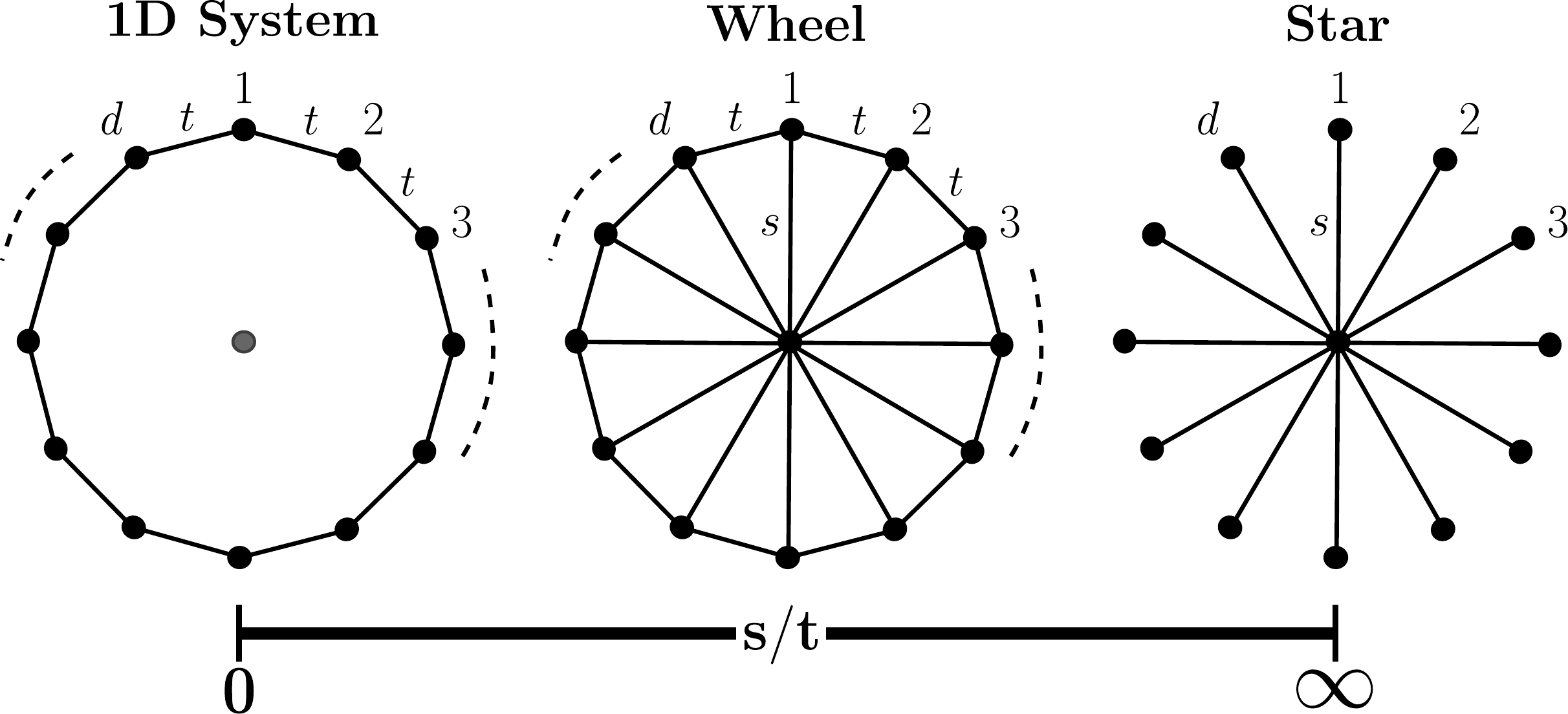}
  \caption{{\bf{Topology of the lattice.}} An interpolation between the $1D$ regime  and the `star`'  through the `wheel' .}
\label{fig:1D_Wheel_Star}
\end{figure}

In contrast to the rather involved experimental realisation of algebraically decaying hopping rates our proposal to realize `infinite'-range hopping will be surprisingly  simple. As it is illustrated in Fig.~\ref{fig:1D_Wheel_Star}, we consider $N$ HCBs on a lattice consisting of a ring with $d$ sites, lattice constant $a$  and \textit{one} additional site at its center. The ring gives rise to a hopping between nearest neighbors at a rate $t >0$. The crucial point is now that the ring's topology allows hopping between the central site and any ring site at a rate $s \geq 0$. Accordingly, the central site has an effect similar to an impurity, making the lattice inhomogeneous.

We remind the reader that  proposing and studying this model shall be seen as one of our key achievements.  It is also worth noticing that various other studies of  BEC for inhomogeneous lattices differ significantly  from ours. They either consider the rather trivial case of ideal bosons \cite{Burioni2000,Burioni2001,Buonsante2002,Brunelli2004,Vidal2011,Oliveira2013,Lyra2014} or restrict to the mean field regime  \cite{Buonsante2004,Halu2012}. At the same time, our model could be particularly appealing to experimentalists since the underlying graph emerges from a  Mexican hat potential(see below) and HCBs can be realized experimentally \cite{DePue1999,Paredes2004} by tuning the  interactions at the Feshbach resonance \cite{Bloch2008, Chin2010, Weidemueller2011,Zuern2012}.

Accordingly, the Hamiltonian of our proposed model of bosons with hard-core interaction reads
\begin{equation}\label{eq:Hamiltonian-wheel}
\hat{H}= -t\sum_{i=1}^{d} [h_{i}^{\dagger} h_{i+1} + h_{i+1}^{\dagger} h_{i}]  -s \sum_{i=1}^{d} [h_{i}^{\dagger} h_{c} + h_{c}^{\dagger} h_{i} ]  \ ,
\end{equation}
where $h_{c}^{\dagger}, h_{c}$ denote the corresponding operators for the central site.
For $s/t \to 0$,  $\hat{H}$ reduces to the pure ring-model (left of Fig.~\ref{fig:1D_Wheel_Star})  and the limit $s/t \to \infty$ leads to the star-model (right of Fig.~\ref{fig:1D_Wheel_Star}).  The solution of the eigenvalue problem for these two limiting cases  is known. For $s=0$ it follows from the solution for impenetrable bosons \cite{Girardeau1960,Lieb1963a,Lieb1963b} which only exhibits quasi-condensation, and  $s= \infty$ was solved in Ref.~\cite{Tennie2017} proving
the existence of true  BEC with  maximal possible number $N_0(N,d)=N(d-N+1)/d$ of condensed bosons.
For finite values of $s/t$ the Hamiltonian (\ref{eq:Hamiltonian-wheel}) interpolates between the ring-lattice and the star-lattice (cf.~Fig.~\ref{fig:1D_Wheel_Star}). Hence, changing the single parameter $s/t$  allows us to investigate in a systematic way the crossover from the regime of quasi-condensation to maximally possible condensation, eventually leading to a number of remarkable insights.

\vspace{0.3cm}
\noindent\textbf{Spectral properties, BEC and entanglement.} The present section contains only the crucial steps. Technical details can be found in the section `Methods' and particularly in the `Supplementary Methods'.

Since the central site couples to the $(N-1)$-
and $N$-particle state space on the ring, a simple and fully analytic solution does not exist. Yet, after implementing a number of steps, the eigenvalue problem for Hamiltonian (\ref{eq:Hamiltonian-wheel}) can be rewritten as

\begin{figure}
 \includegraphics[width=7cm]{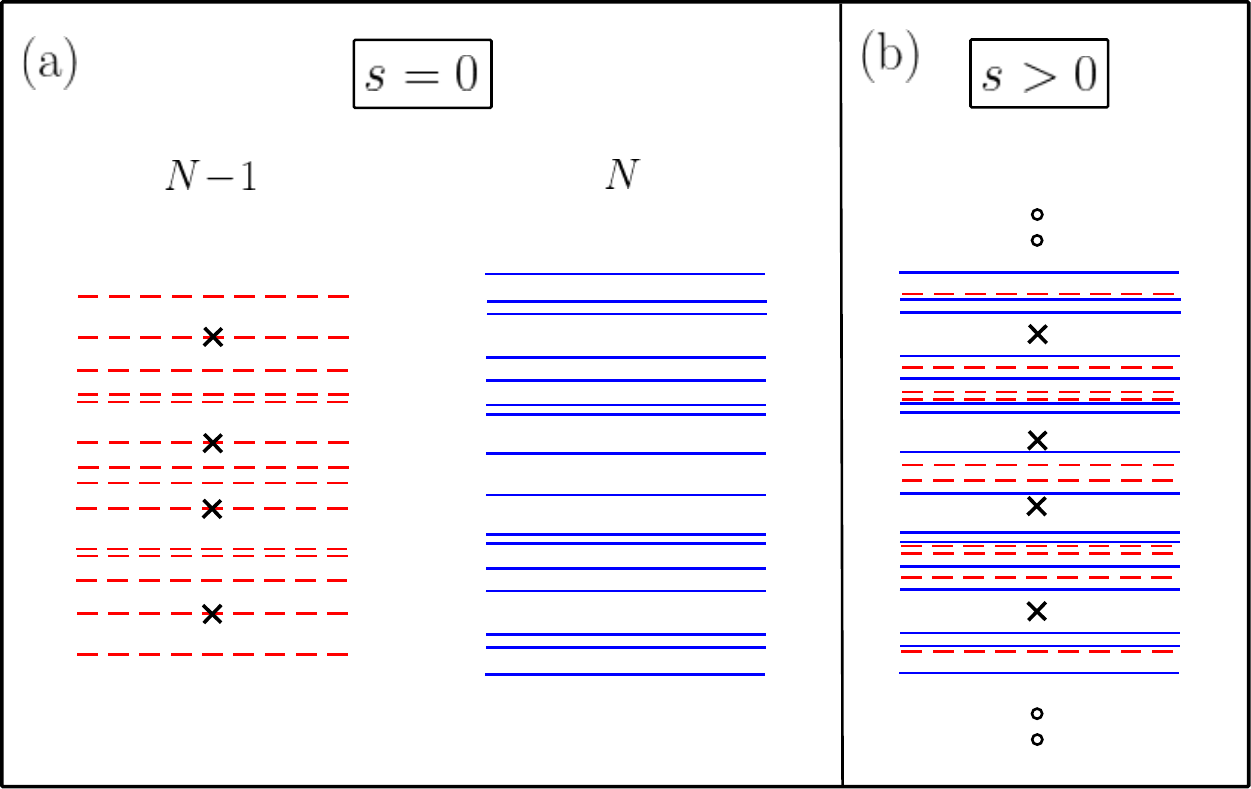}
  \caption{{\bf{Schematic representation of the unperturbed and perturbed spectrum for ${\bf{n \leq 1/2}}$.}}  (a) The unperturbed band spectrum of $(N-1)$ hard-core bosons (red dashed lines) and $N$ hard-core bosons (blue solid lines). The crosses mark those levels which disappear under the perturbation.  (b) The perturbed spectrum consisting of the shifted unperturbed levels(dashed red and solid blue lines) and the crosses mark those levels of the unperturbed $(N-1)$-particle spectrum which have disappeared. The open circles  are the new levels  generated by the perturbation. }
\label{fig:spectra}
\end{figure}

\begin{equation}\label{eq:spectrum-1}
1=s^2F^{(N)}_d(E;\{A_{\boldsymbol{\nu}}\})   \ ,
\end{equation}
where $E$ is the eigenvalue and  $\{A_{\boldsymbol{\nu}}\}$ are amplitudes of the unperturbed (i.e., corresponding to $s=0$) $N$-particle eigenstates $\ket{\psi^{0}_{\boldsymbol{\nu}}(N)}$ on the ring. Although this equation cannot be solved analytically for the entire regime of $s$, it allows us to derive in a nontrivial way  important qualitatively correct features of the spectrum.
The unperturbed $(N-1)$- and $N$-particle spectrum forms a band of discrete levels (see Fig.~2a) which becomes continuous for $d \to \infty$. The hopping between the central site and the ring introduces a `hybridization' of these two spectra leading on one hand to a shift of order $1/d$ of the unperturbed band-levels. On the other hand, some energy levels (marked by crosses) of the smaller $(N-1)$-particle band(assuming $n=N/d < 1/2$, which is not a restriction due to the particle-hole duality) are found to disappear. These levels, however, reappear as new discrete eigenvalues symmetrically below and above the perturbed $N$-particle band(see open circles in Fig.~2b).  The larger $s$ and $N$, the more of those new discrete energy levels occur. As a matter of fact, they follow from the eigenvalues of an effective Hamiltonian for $N$ HCBs with `infinite'-range hopping: $\hat{H}^{\mathrm{eff}}= \tilde{s}^2 (1/d) \sum^{d}_{i,j=1} h^{\dagger}_i h_j$. Here, the parameter $\tilde{s}=(s/t) \sqrt{d}$ is a scaled dimensionless hopping rate.
This mapping of the original model to an effective one holds for  $\tilde{s} \gg  2\sqrt{2} \ \pi /\sqrt{d}$ for the diluted gas($n \ll 1$) and in case of finite $n$  for  $\tilde{s} \gg (4/\pi)\sqrt{d} \sin{\pi n}/{\sqrt{n(1-n})}$.

Most importantly, these findings imply also the opening of an energy gap $\Delta E = E^0_{\mathrm{low}} - E_0$ between the perturbed ground state energy $E_0$ and  $E^0_{\mathrm{low}}$, the lower edge of the $N$-particle band:
\begin{equation}\label{eq:eigenvalue}
\begin{split}
\Delta E \simeq  |E_F/2| \big[\sqrt{(1 + \tilde{s}^2 d\ n(1-n)/ (E_F/2t)^2}  -1 \big]   \  \  .
\end{split}
\end{equation}
Also the number $N_0$ of condensed HCBs can be derived analytically since it is related to the largest eigenvalue of $H^{\mathrm{eff}}$. We obtain
\begin{equation}\label{eq:condensate-number}
N_0  \simeq  N \big[(1-n) -  |\beta|^2(1-2 n) N^{-1} \big]  \  ,
\end{equation}
where the prefactor $ |\beta|^2$ of the $1/N$-correction is given in the Supplementary Eq.~S30.

\begin{figure}
\includegraphics[width=7cm]{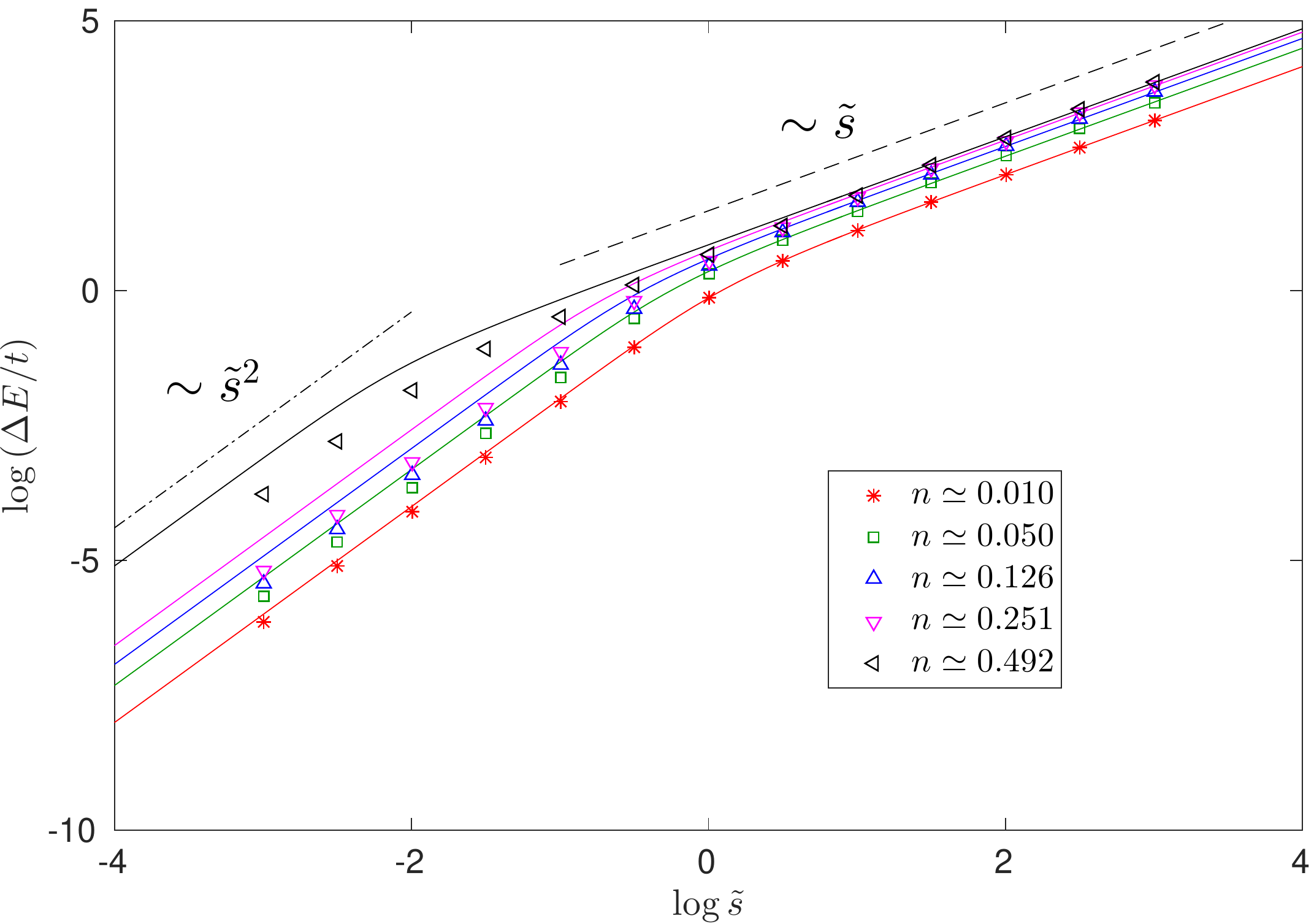}
\caption{{\bf{Excitation gap.}} log-log plot for the excitation gap as a function of $\tilde{s}$ for $d=199$ and various filling factors $n$. Results from density matrix renormalization group calculations(symbols) and analytical result(Eq.(\ref{eq:eigenvalue}))(solid lines). Dashed-dotted and dashed line represent slope two and one, respectively.}
\label{fig:energy-gap}
\end{figure}

\begin{figure}
\includegraphics[width=7.0cm]{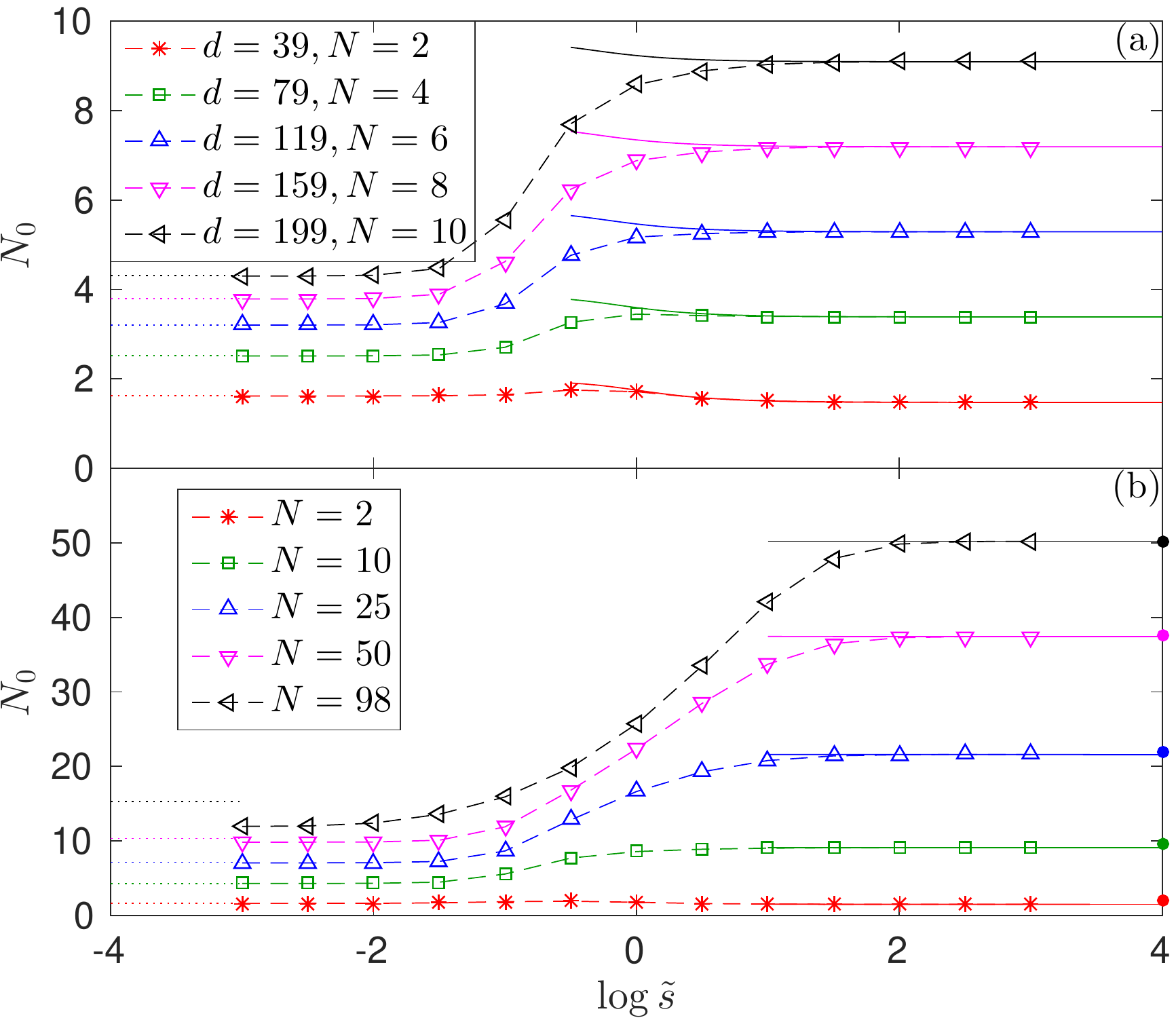}
 \caption[DMRG result]{{\bf{Number of condensed bosons.}} (a) $N_0$ as a function of $\log(\tilde{s})$ for  fixed low density $n \simeq 0.05$ and various site numbers $d$. (b) $N_0$ as a function of $\log(\tilde{s})$, various numbers of particles $N$, and fixed number of sites $d=199$. The dots on the vertical axis represent  $N(1-n)$.
 Symbols: Results from density matrix renormalization group calculations, dashed lines: guide for the eye, solid lines: analytical result (Supplementary equation S28). The dotted lines mark the asymptotic values for finite $N$, $s=0, d \to \infty$ obtained by the exact numerical calculation of a Toeplitz determinant \cite{Forrester2003b}.}
 \label{condensate-number}
\end{figure}

In order to support these analytical results and to extend those for finite $d$ to small and intermediate values of $\tilde{s}$  we have performed large-scale density matrix renormalization group computations (DMRG) \cite{White1992b,White1993,Schollwock2005}. The corresponding  results together with  the analytical ones are presented in  Figs.~\ref{fig:energy-gap} and  \ref{condensate-number}.
The log-log representation of the gap $\Delta E(\tilde{s})$ in Fig.~\ref{fig:energy-gap} reveals a distinctive crossover from a $\tilde{s}^2$-dependence for  $\tilde{s} \ll 1$ to the linear dependence on $\tilde{s}$ for $\tilde{s} \gg 1$. For the diluted gas, i.e.~$n \ll 1$, the analytical and DMRG  results in the $\tilde{s}^2$- and $\tilde{s}$-regime are in  good agreement.  When the density is increased this agreement remains excellent in the linear regime while it gets worse in the complementary range. Fig.~\ref{condensate-number} illustrates clearly for the diluted gas (Fig. 4a) and for higher densities (Fig. 4b)  the crossover from a quasi-condensate with $N_0 \sim \sqrt{N}$ to the maximally possible condensation  $N_0(N,n) \simeq N(1-n)$. The deviations from the $\sqrt{N}$-dependence for small $\tilde{s}$ and higher densities (see lower panel) is an effect of the lattice-discreteness. In the regime in which the mapping to the effective Hamiltonian is valid (see above)  the analytical and DMRG results agree well.

\begin{figure}[htb]
 \includegraphics[width=7cm]{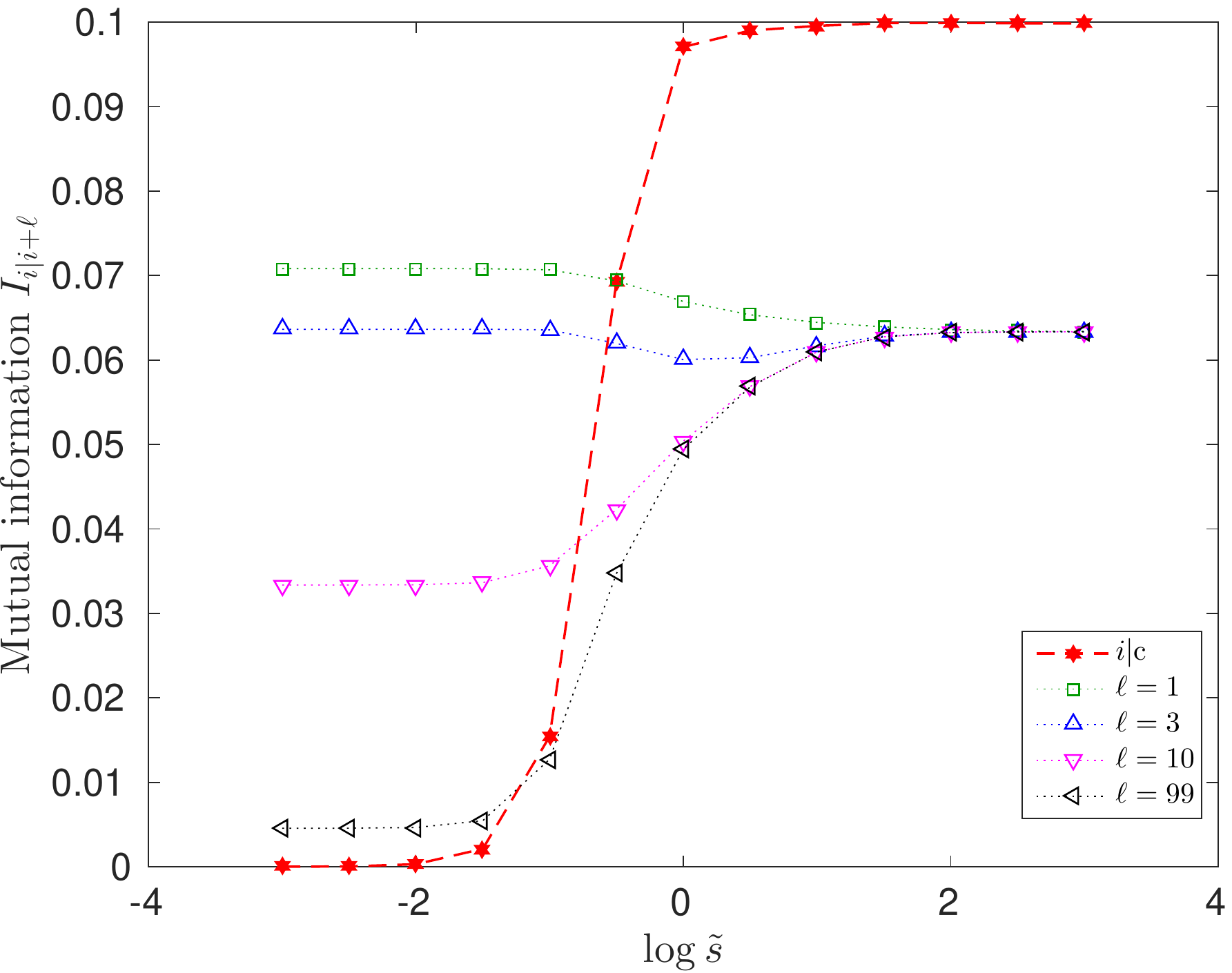}
\caption{{\bf{Quantum informational quantity.}} Two-site correlation measured in terms of the mutual information $I_{i|\mathrm{c}}$ between the central site and a ring site
(red stars), and $I_{i|i+\ell}$ for two sites on the ring separated by distance $\ell=1,3,10,99$
for $d=199$ and $n\simeq 0.05$(other symbols). The dashed and dotted lines are a guide for the eye.}
\label{fig:mutual_information}
\end{figure}

To explore a possible relation between  BEC and the entanglement structure of the ground state we have used DMRG for calculating the mutual information  between the central site $c$ and any ring site $i$  ($I_{i|c}$) and between two $\ell$-th nearest neighbor ring sites ($I_{i|i+\ell}$)(see Methods). The corresponding results for $d=199$ and $n\simeq 0.05$ are shown in Fig.~\ref{fig:mutual_information}.
The change in the respective pattern related to the crossover from quasi-BEC to genuine BEC is clearly visible through the mutual information, as well. The correlation between the central and any ring site, $I_{i|c}$, vanishes for $\tilde{s}$ small while it saturates to a finite value in the limit of large $\tilde{s}$ when the model exhibits `infinite'-range hopping. $I_{i|i+l}$ saturates  also with increasing $\tilde{s}$ to a constant value for all $\ell$ demonstrating the  growth of long range correlations. This  relates to the generation of BEC. For $\tilde{s}=0$,  $I_{i|i+\ell}$ decays algebraically  with increasing $\ell$ which reflects the algebraic dependence of the quasi-condensate on $N$. Whereas for finite values of  $\tilde{s}$ its decay becomes exponential as the gap opens, and saturates to finite value for very large $\ell$ values.

\begin{figure}
 \includegraphics[width=8.4cm]{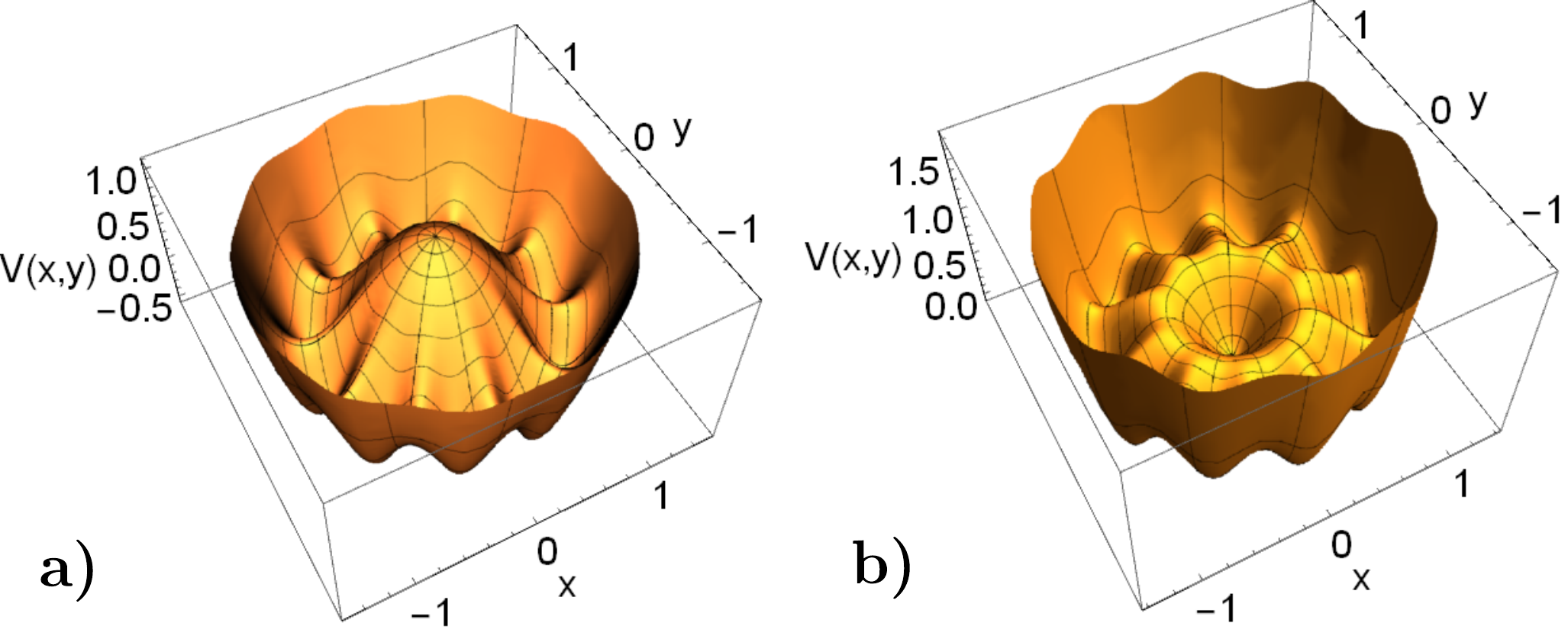}
 \caption[dipped Mexican head.]{{\bf{Mexican-hat-type trap potential.}} (a) Realization of the ring lattice for  $d=10$ by a Mexican-hat-potential. (b) Realization of the  wheel lattice  for $d=10$ by a Mexican-hat-like potential
 with a local well at its center. Loading hard-core bosons  into the potential landscape in (a) and creating a local well as shown in (b) generates a crossover from quasi to complete Bose-Einstein condensation.}
 \label{mexican-hat(lattice)-1}
\end{figure}

\vspace{0.3cm}
\noindent\textbf{Potential experimental realization.} As a possible experimental realization of our model (\ref{eq:Hamiltonian-wheel}) we propose in a first step to confine $N$ ultracold bosonic atoms into two dimensions subject to  a Mexican-hat-type potential $V(x,y)$ with $d$ local wells (Fig.~\ref{mexican-hat(lattice)-1}a) in complete analogy to several recent years' experiments \cite{Amico2005,Franke2007,Ramanathan2011,Amico2014,Bell2016}. Then, one may tune the interaction at the Feshbach resonance to realize HCBs in the same way as reported in Ref.~\cite{Paredes2004} for a cigar-shaped confinement to realize quasi-condensation of HCB with $N_0(N) \propto N^{1/2}$. Next, creation of a local well at the hat's center (Fig.~\ref{mexican-hat(lattice)-1}b) and increasing its depth more and more would strongly enhance the mobility of the HCBs due to their possible transitions back and forth between any ring-well and the central one. This would significantly change the physical behavior and BEC would occur with $N_0(N) \sim N$. In order for this to happen already for \textit{finite} $d$ it must be  $s/t \gg 2\sqrt{2}\pi/d$ in case of a diluted gas (see previous section) which is the regime relevant for ultracold gases. The hopping occurs due to tunneling between the corresponding wells. Let $(V_r, l_r=a)$ and  $(V_c, l_c=ad/(2\pi))$ denote the potential barrier and tunneling distance, respectively,  between two adjacent ring-wells and between a ring-well and the central one.  Use of the WKB tunneling rate yields the estimate $s/t \approx (\gamma_c/\gamma_r) \exp[-\sqrt{ma^2/\hbar^2}(\sqrt{V_c}d/(2\pi) -\sqrt{V_r})]$ with $m$ the particle's mass and $\gamma_{\alpha}, \alpha=c,r$  the so-called attempt frequency related to the zero-point oscillation frequency in the corresponding well. For instance, if $d=79$ and $N=4$
(one data set in Fig.~\ref{condensate-number}(a)) `BEC'-like behavior should occur for $s/t > 1$. This can be satisfied
if $V_c/V_r \approx (2\pi/d)^2$ or if $a$ compared to $\hbar/\sqrt{m \ \max\{V_c,V_r\}}$ is small enough,  provided $\gamma_c/\gamma_r \approx 1$.

If the trap potential in Fig.~\ref{mexican-hat(lattice)-1} is chosen such that it represents a good experimental realization of the `wheel' lattice (cf. Fig.~\ref{fig:1D_Wheel_Star}) there is true condensation for sufficiently large $s/t$. In particular, since only a single one-particle state (zero-momentum state) is macroscopically occupied, no fragmented condensation exists per definition. This is consistent with the expectation that homogeneous bosonic systems with purely repelling pair interactions do not exhibit fragmented condensation  \cite{Pitaevskii2016}.  Although the presence of the central well(central site) makes the system inhomogeneous it can not generate fragmentation, because it  accommodates  maximally one HCB, only.
But increasing the width of the central well in Fig.~\ref{mexican-hat(lattice)-1}b such that  it can accommodate a ``macroscopic number of  bosons of an ultracold gas, a situation similar to the double-well-like trap potential in one dimension occurs \cite{Sakmann2008}. As shown in that work, fragmented condensation may then occur if the barrier height of the double well is high enough.

It is worth noticing that according to the DMRG results (see also Fig.~\ref{condensate-number}) one would not need to realize a macroscopically large ring to observe our crossover. Yet, in case experimentalists could even realise our model with a huge number $d$ of sites on a ring of fixed size (i.e., the limit $d \to \infty, a \to 0$ with $ad$ fixed) this would generate a \textit{true} Mexican-hat potential with \textit{continuous} rotational invariance and the HCBs would become a Tonks-Girardeau gas. Again, creating a central well would generate genuine Bose-Einstein condensation.

\vspace{0.3cm}
\noindent\textbf{\large{Discussions}}\\
We proposed and comprehensively studied a physical model of strongly interacting bosons which allows one to drive a non-trivial transition from quasi-condensation to maximal Bose-Einstein condensation(BEC). It is particularly appealing that this necessitates the tuning of just a single control parameter which changes the underlying topology in such a distinctive way that the `infinite' range hopping model is simulated.
 The enhanced mobility of the bosons then compensates for the destructive effects of the strong interaction to generate BEC.
Without  solving  the model's eigenvalue equation exactly, our  kind of analytical approach(see the  section ``Spectral properties, BEC and entanglement'' above and also the `Supplementary Methods' ) allows us to show on a qualitative level why an excitation gap occurs in the $N$-particle spectrum , which usually is highly demanding. Similarly to, e.g., superconductivity, the quantum Hall effect and the Haldane phase the existence of such a  gap has an enormous influence on the physical behavior, e.g., making the BE-condensate robust to thermal noise and perturbations in general.

It is worth highlighting the striking potential of our mechanism for generating BEC. As a matter of fact, it is conceptually quite different to the
well-known generation of BEC at finite temperatures for non-interacting bosons.
The latter is either merely due to the opening of a  gap in the ``one-particle'' spectrum  or a deformation of the density of states (in analogy to the transition from $D \leq 2$ to $D=3$) \cite{Burioni2000,Burioni2001,Buonsante2002,Brunelli2004,Vidal2011,Oliveira2013,Lyra2014}. The same effectively applies to the experimental \cite{Greiner2001,Dettmer2001,Goerlitz2001,Orzel2001,Paredes2004,Kinoshita2004,Stoeferle2004} and theoretical studies \cite{Das2002,Salasnich2004, Salasnich2005} in which the cylindrical or torroidal confinement is relaxed to reach the mean-field regime. In our system, however, it is the interplay between mobility and interaction within the ``non-perturbative'' regime which generates genuine BEC (see the rather involved derivation in the Supplementary Methods). The non-trivial influence of the interaction is also well-illustrated by the analytical result for the ground state gap (Eq.~(\ref{eq:eigenvalue})) which in the regime of maximal BEC differs from the one of non-interacting bosons by the crucial factor $\sqrt{1-n}$. Remarkably, $1-n$ is nothing else than the universal reduction of the maximal possible degree of condensation due to the hard-core constraint\cite{Tennie2017}, which is the quantum depletion. In case of \textit{finite} on-site interactions, this depletion factor $\nu(n)$ is expected to interpolate between both extremal cases of hard-core and ideal bosons, $1-n \leq \nu(n) \leq 1$. This would provide a remarkable exact relation between the ground state gap, quantum depletion and the interaction strength of the ultracold atoms. Since the latter can systematically be tuned at the Feshbach resonance \cite{Bloch2008, Chin2010, Weidemueller2011,Zuern2012} this would open an  avenue for steering ground state gaps and controlling the number of bosons in BEC.

Finally, inspired by the fruitful interplay of theory and experiments in the field of ultracold gases our work based on analytical and exact large scale DMRG calculations shall be understood as a proposal to the experimentalists as well. Our model could be particularly appealing since the underlying graph emerges from a  Mexican-hat-type potential and the entire transition can be driven by tuning just a single
control parameter. It is then exactly the respective central site which can be probed to confirm that transition. At the same time, this would also  exploit the fruitful link\cite{Osterloh2002,Osborne2002} between quantum phase transitions and entanglement or related promising quantum informational theoretical concepts, as illustrated in  Fig.~\ref{fig:mutual_information}.
\\
\\

\vspace{0.3cm}
\noindent\textbf{\large{Methods}}\\
\noindent\textbf{Eigenvalue problem.} The central site generates for the $N$-particle state a superposition $\ket{\Psi_N} = \alpha \ket{\phi_N}_{r}\otimes \ket{0}_c + \beta \ket{\varphi_{N-1}}_{r}\otimes \ket{1}_c$ of a $N$- and $(N-1)$-particle
ring-state. Expansion of these states with respect to the unperturbed $(N-1)$- and $N$-particle
ring-states allows to decouple the original eigenvalue problem. This leads to a nonintegrable eigenvalue problem on the ring itself. Straightforward manipulation allows to derive Eq.~(\ref{eq:spectrum-1}). For details see the Supplementary Methods.

\vspace{0.3cm}
\noindent\textbf{Density matrix renormalization group.} The DMRG calculations were performed for $d \leq 199$  and $N \leq 98$.
In the DMRG procedure we have performed calculations using the dynamic block state selection approach \cite{Legeza-2004b}.
We have set a tight error bound on the diagonalization procedure, i.e., we set the
residual error of the Davidson method to $10^{-9}$ and used ten DMRG sweeps.
We have checked that the various quantities of interest are practically insensitive on the bond dimension being larger than 1024.

Besides calculating energy eigenvalues and the one-($\rho_i$) and two-site($\rho_{ij}$) reduced
density matrices we have also determined one- and two-site von Neumann
entropies $s_i$ and $s_{ij}$, respectively, as well as the two-site mutual
information, $I_{i|j}$, given as $I_{i|j}=s_i+s_j-s_{ij}$\cite{Rissler2006,Szalay2015}.
Here $s_i=-{\rm Tr}\rho_i\ln\rho_i$ and $s_{ij}=-{\rm Tr}\rho_{ij}\ln\rho_{ij}$.
\vspace{0.3cm}

\vspace{0.3cm}
\noindent\textbf{\large{Acknowledgements}}\\
We gratefully acknowledge critical comments on the present manuscript by P.\hspace{0.5mm}van Dongen, F.\hspace{0.5mm}Gebhard, J.\hspace{0.5mm}Marino and  L.\hspace{0.5mm}Pollet.
We also would like to thank P.J.\hspace{0.5mm}Forrester and T.M.\hspace{0.5mm}Garoni for providing the exact results obtained  directly from the Toeplitz determinant for the number of condensed particles in one dimension for small $N$.
This work has been supported in part by the Hungarian  National  Research,  Development  and
Innovation  Office  (grant  no. K120569 and K134983), and  the  Hungarian  Quantum  Technology
National  Excellence  Program  (project no. 2017-1.2.1-NKP-2017-00001).
{\"O}.L. acknowledges financial support from the Alexander von Humboldt foundation.
M.M. has been supported by the \'UNKP-19-3 Hungarian New National Excellence Program of the Ministry for Innovation and Technology.
The development of the DMRG libraries was supported by the Center for
Scalable and Predictive methods for Excitation and Correlated phenomena
(SPEC), which is funded from the Computational Chemical Sciences Program by
the U.S. Department of Energy (DOE), at Pacific Northwest National
Laboratory.
C.S. acknowledges financial support from the UK Engineering and Physical Sciences Research Council (Grant EP/P007155/1) and  Deutsche Forschungsgemeinschaft (Grant SCHI 1476/1-1).
\vspace{0.7cm}

\bibliography{Ref_HCB2}

\begin{thebibliography}{92}%
\makeatletter
\providecommand \@ifxundefined [1]{%
 \@ifx{#1\undefined}
}%
\providecommand \@ifnum [1]{%
 \ifnum #1\expandafter \@firstoftwo
 \else \expandafter \@secondoftwo
 \fi
}%
\providecommand \@ifx [1]{%
 \ifx #1\expandafter \@firstoftwo
 \else \expandafter \@secondoftwo
 \fi
}%
\providecommand \natexlab [1]{#1}%
\providecommand \enquote  [1]{``#1''}%
\providecommand \bibnamefont  [1]{#1}%
\providecommand \bibfnamefont [1]{#1}%
\providecommand \citenamefont [1]{#1}%
\providecommand \href@noop [0]{\@secondoftwo}%
\providecommand \href [0]{\begingroup \@sanitize@url \@href}%
\providecommand \@href[1]{\@@startlink{#1}\@@href}%
\providecommand \@@href[1]{\endgroup#1\@@endlink}%
\providecommand \@sanitize@url [0]{\catcode `\\12\catcode `\$12\catcode
  `\&12\catcode `\#12\catcode `\^12\catcode `\_12\catcode `\%12\relax}%
\providecommand \@@startlink[1]{}%
\providecommand \@@endlink[0]{}%
\providecommand \url  [0]{\begingroup\@sanitize@url \@url }%
\providecommand \@url [1]{\endgroup\@href {#1}{\urlprefix }}%
\providecommand \urlprefix  [0]{URL }%
\providecommand \Eprint [0]{\href }%
\providecommand \doibase [0]{https://doi.org/}%
\providecommand \selectlanguage [0]{\@gobble}%
\providecommand \bibinfo  [0]{\@secondoftwo}%
\providecommand \bibfield  [0]{\@secondoftwo}%
\providecommand \translation [1]{[#1]}%
\providecommand \BibitemOpen [0]{}%
\providecommand \bibitemStop [0]{}%
\providecommand \bibitemNoStop [0]{.\EOS\space}%
\providecommand \EOS [0]{\spacefactor3000\relax}%
\providecommand \BibitemShut  [1]{\csname bibitem#1\endcsname}%
\let\auto@bib@innerbib\@empty
\bibitem [{\citenamefont {Griffin}\ \emph {et~al.}(1995)\citenamefont
  {Griffin}, \citenamefont {Snoke},\ and\ \citenamefont
  {Stringari}}]{Griffin1995}%
  \BibitemOpen
  \bibfield  {author} {\bibinfo {author} {\bibfnamefont {A.}~\bibnamefont
  {Griffin}}, \bibinfo {author} {\bibfnamefont {D.~W.}\ \bibnamefont {Snoke}},\
  and\ \bibinfo {author} {\bibfnamefont {S.}~\bibnamefont {Stringari}},\
  }\href@noop {} {\emph {\bibinfo {title} {Bose-Einstein Condensation}}}\
  (\bibinfo  {publisher} {Cambridge University Press, Cambridge},\ \bibinfo
  {year} {1995})\BibitemShut {NoStop}%
\bibitem [{\citenamefont {Pethick}\ and\ \citenamefont
  {Smith}(2002)}]{Pethick2002}%
  \BibitemOpen
  \bibfield  {author} {\bibinfo {author} {\bibfnamefont {C.}~\bibnamefont
  {Pethick}}\ and\ \bibinfo {author} {\bibfnamefont {H.}~\bibnamefont
  {Smith}},\ }\href@noop {} {\emph {\bibinfo {title} {Bose-{E}instein
  {C}ondensation in {D}ilute {G}ases}}}\ (\bibinfo  {publisher} {Cambridge
  University Press, Cambridge},\ \bibinfo {year} {2002})\BibitemShut {NoStop}%
\bibitem [{\citenamefont {Leggett}(2006)}]{Leggett2006}%
  \BibitemOpen
  \bibfield  {author} {\bibinfo {author} {\bibfnamefont {A.}~\bibnamefont
  {Leggett}},\ }\href@noop {} {\emph {\bibinfo {title} {Quantum {L}iquid: Bose
  {C}ondensation and {C}ooper {P}airing in {C}ondensed-{M}atter {S}ystems}}}\
  (\bibinfo  {publisher} {Oxford University Press, Oxford},\ \bibinfo {year}
  {2006})\BibitemShut {NoStop}%
\bibitem [{\citenamefont {Pitaevskii}\ and\ \citenamefont
  {Stringari}(2016)}]{Pitaevskii2016}%
  \BibitemOpen
  \bibfield  {author} {\bibinfo {author} {\bibfnamefont {L.~P.}\ \bibnamefont
  {Pitaevskii}}\ and\ \bibinfo {author} {\bibfnamefont {S.}~\bibnamefont
  {Stringari}},\ }\href@noop {} {\emph {\bibinfo {title} {Bose-Einstein
  Condensation and Superfluidity}}}\ (\bibinfo  {publisher} {Oxford University
  Press, Oxford},\ \bibinfo {year} {2016})\BibitemShut {NoStop}%
\bibitem [{\citenamefont {Anderson}\ \emph {et~al.}(1995)\citenamefont
  {Anderson}, \citenamefont {Ensher}, \citenamefont {Matthews}, \citenamefont
  {Wieman},\ and\ \citenamefont {Cornell}}]{Anderson1995}%
  \BibitemOpen
  \bibfield  {author} {\bibinfo {author} {\bibfnamefont {M.~H.}\ \bibnamefont
  {Anderson}}, \bibinfo {author} {\bibfnamefont {J.~R.}\ \bibnamefont
  {Ensher}}, \bibinfo {author} {\bibfnamefont {M.~R.}\ \bibnamefont
  {Matthews}}, \bibinfo {author} {\bibfnamefont {C.~E.}\ \bibnamefont
  {Wieman}},\ and\ \bibinfo {author} {\bibfnamefont {E.~A.}\ \bibnamefont
  {Cornell}},\ }\bibfield  {title} {\bibinfo {title} {Observation of
  {B}ose-{E}instein condensation in a dilute atomic vapor},\ }\href
  {https://doi.org/10.1126/science.269.5221.198} {\bibfield  {journal}
  {\bibinfo  {journal} {Science}\ }\textbf {\bibinfo {volume} {269}},\ \bibinfo
  {pages} {198} (\bibinfo {year} {1995})}\BibitemShut {NoStop}%
\bibitem [{\citenamefont {Bradley}\ \emph {et~al.}(1995)\citenamefont
  {Bradley}, \citenamefont {Sackett}, \citenamefont {Tollett},\ and\
  \citenamefont {Hulet}}]{Bradley1995}%
  \BibitemOpen
  \bibfield  {author} {\bibinfo {author} {\bibfnamefont {C.~C.}\ \bibnamefont
  {Bradley}}, \bibinfo {author} {\bibfnamefont {C.~A.}\ \bibnamefont
  {Sackett}}, \bibinfo {author} {\bibfnamefont {J.~J.}\ \bibnamefont
  {Tollett}},\ and\ \bibinfo {author} {\bibfnamefont {R.~G.}\ \bibnamefont
  {Hulet}},\ }\bibfield  {title} {\bibinfo {title} {Evidence of
  {B}ose-{E}instein {C}ondensation in an {A}tomic {G}as with {A}ttractive
  {I}nteractions},\ }\href {https://doi.org/10.1103/PhysRevLett.75.1687}
  {\bibfield  {journal} {\bibinfo  {journal} {Phys. Rev. Lett.}\ }\textbf
  {\bibinfo {volume} {75}},\ \bibinfo {pages} {1687} (\bibinfo {year}
  {1995})}\BibitemShut {NoStop}%
\bibitem [{\citenamefont {Davis}\ \emph {et~al.}(1995)\citenamefont {Davis},
  \citenamefont {Mewes}, \citenamefont {Andrews}, \citenamefont {van Druten},
  \citenamefont {Durfee}, \citenamefont {Kurn},\ and\ \citenamefont
  {Ketterle}}]{Davis1995}%
  \BibitemOpen
  \bibfield  {author} {\bibinfo {author} {\bibfnamefont {K.~B.}\ \bibnamefont
  {Davis}}, \bibinfo {author} {\bibfnamefont {M.-O.}\ \bibnamefont {Mewes}},
  \bibinfo {author} {\bibfnamefont {M.~R.}\ \bibnamefont {Andrews}}, \bibinfo
  {author} {\bibfnamefont {N.~J.}\ \bibnamefont {van Druten}}, \bibinfo
  {author} {\bibfnamefont {D.~S.}\ \bibnamefont {Durfee}}, \bibinfo {author}
  {\bibfnamefont {D.~M.}\ \bibnamefont {Kurn}},\ and\ \bibinfo {author}
  {\bibfnamefont {W.}~\bibnamefont {Ketterle}},\ }\bibfield  {title} {\bibinfo
  {title} {Bose-{E}instein condensation in a gas of {S}odium atoms},\ }\href
  {https://doi.org/10.1103/PhysRevLett.75.3969} {\bibfield  {journal} {\bibinfo
   {journal} {Phys. Rev. Lett.}\ }\textbf {\bibinfo {volume} {75}},\ \bibinfo
  {pages} {3969} (\bibinfo {year} {1995})}\BibitemShut {NoStop}%
\bibitem [{\citenamefont {Bogoliubov}(1947)}]{Bogoliubov1947}%
  \BibitemOpen
  \bibfield  {author} {\bibinfo {author} {\bibfnamefont {N.~N.}\ \bibnamefont
  {Bogoliubov}},\ }\bibfield  {title} {\bibinfo {title} {On the theory of
  superfluidity},\ }\href {https://doi.org/10.1016/B978-0-08-015816-7.50020-1}
  {\bibfield  {journal} {\bibinfo  {journal} {J. Phys. USSR}\ }\textbf
  {\bibinfo {volume} {11}},\ \bibinfo {pages} {23} (\bibinfo {year}
  {1947})}\BibitemShut {NoStop}%
\bibitem [{\citenamefont {Gross}(1961)}]{Gross1961}%
  \BibitemOpen
  \bibfield  {author} {\bibinfo {author} {\bibfnamefont {E.}~\bibnamefont
  {Gross}},\ }\bibfield  {title} {\bibinfo {title} {Structure of a quantized
  vortex in boson systems},\ }\href
  {https://link.springer.com/article/10.1007/BF02731494} {\bibfield  {journal}
  {\bibinfo  {journal} {Nuovo Cimento}\ }\textbf {\bibinfo {volume} {20}},\
  \bibinfo {pages} {454} (\bibinfo {year} {1961})}\BibitemShut {NoStop}%
\bibitem [{\citenamefont {Gross}(1963)}]{Gross1963}%
  \BibitemOpen
  \bibfield  {author} {\bibinfo {author} {\bibfnamefont {E.}~\bibnamefont
  {Gross}},\ }\bibfield  {title} {\bibinfo {title} {Hydrodynamics of a
  superfluid condensate},\ }\href {https://doi.org/10.1063/1.1703944}
  {\bibfield  {journal} {\bibinfo  {journal} {J. Math. Phys.}\ }\textbf
  {\bibinfo {volume} {4}},\ \bibinfo {pages} {195} (\bibinfo {year}
  {1963})}\BibitemShut {NoStop}%
\bibitem [{\citenamefont {P.}(1961)}]{Pitaevskii1961}%
  \BibitemOpen
  \bibfield  {author} {\bibinfo {author} {\bibfnamefont {P.~L.}\ \bibnamefont
  {P.}},\ }\bibfield  {title} {\bibinfo {title} {Vortex lines in an imperfect
  bose gas},\ }\href@noop {} {\bibfield  {journal} {\bibinfo  {journal} {Sov.
  Phys. JETP}\ }\textbf {\bibinfo {volume} {13}},\ \bibinfo {pages} {451}
  (\bibinfo {year} {1961})}\BibitemShut {NoStop}%
\bibitem [{\citenamefont {Lee}\ and\ \citenamefont {Yang}(1957)}]{Lee1957a}%
  \BibitemOpen
  \bibfield  {author} {\bibinfo {author} {\bibfnamefont {T.~D.}\ \bibnamefont
  {Lee}}\ and\ \bibinfo {author} {\bibfnamefont {C.~N.}\ \bibnamefont {Yang}},\
  }\bibfield  {title} {\bibinfo {title} {Many-{B}ody {P}roblem in {Q}uantum
  {M}echanics and {Q}uantum {S}tatistical {M}echanics},\ }\href
  {https://doi.org/10.1103/PhysRev.105.1119} {\bibfield  {journal} {\bibinfo
  {journal} {Phys. Rev.}\ }\textbf {\bibinfo {volume} {105}},\ \bibinfo {pages}
  {1119} (\bibinfo {year} {1957})}\BibitemShut {NoStop}%
\bibitem [{\citenamefont {Lee}\ \emph {et~al.}(1957)\citenamefont {Lee},
  \citenamefont {Huang},\ and\ \citenamefont {Yang}}]{Lee1957b}%
  \BibitemOpen
  \bibfield  {author} {\bibinfo {author} {\bibfnamefont {T.~D.}\ \bibnamefont
  {Lee}}, \bibinfo {author} {\bibfnamefont {K.}~\bibnamefont {Huang}},\ and\
  \bibinfo {author} {\bibfnamefont {C.~N.}\ \bibnamefont {Yang}},\ }\bibfield
  {title} {\bibinfo {title} {Eigenvalues and {E}igenfunctions of a {B}ose
  {S}ystem of {H}ard {S}pheres and {I}ts {L}ow-{T}emperature {P}roperties},\
  }\href {https://doi.org/10.1103/PhysRev.106.1135} {\bibfield  {journal}
  {\bibinfo  {journal} {Phys. Rev.}\ }\textbf {\bibinfo {volume} {106}},\
  \bibinfo {pages} {1135} (\bibinfo {year} {1957})}\BibitemShut {NoStop}%
\bibitem [{\citenamefont {Brueckner}\ and\ \citenamefont
  {Sawada}(1957{\natexlab{a}})}]{Brueckner1957a}%
  \BibitemOpen
  \bibfield  {author} {\bibinfo {author} {\bibfnamefont {K.~A.}\ \bibnamefont
  {Brueckner}}\ and\ \bibinfo {author} {\bibfnamefont {K.}~\bibnamefont
  {Sawada}},\ }\bibfield  {title} {\bibinfo {title} {Bose-{E}instein {G}as with
  {R}epulsive {I}nteractions: {G}eneral {T}heory},\ }\href
  {https://doi.org/10.1103/PhysRev.106.1117} {\bibfield  {journal} {\bibinfo
  {journal} {Phys. Rev.}\ }\textbf {\bibinfo {volume} {106}},\ \bibinfo {pages}
  {1117} (\bibinfo {year} {1957}{\natexlab{a}})}\BibitemShut {NoStop}%
\bibitem [{\citenamefont {Brueckner}\ and\ \citenamefont
  {Sawada}(1957{\natexlab{b}})}]{Brueckner1957b}%
  \BibitemOpen
  \bibfield  {author} {\bibinfo {author} {\bibfnamefont {K.~A.}\ \bibnamefont
  {Brueckner}}\ and\ \bibinfo {author} {\bibfnamefont {K.}~\bibnamefont
  {Sawada}},\ }\bibfield  {title} {\bibinfo {title} {Bose-{E}instein {G}as with
  {R}epulsive {I}nteractions: {H}ard {S}pheres at {H}igh {D}ensity},\ }\href
  {https://doi.org/10.1103/PhysRev.106.1128} {\bibfield  {journal} {\bibinfo
  {journal} {Phys. Rev.}\ }\textbf {\bibinfo {volume} {106}},\ \bibinfo {pages}
  {1128} (\bibinfo {year} {1957}{\natexlab{b}})}\BibitemShut {NoStop}%
\bibitem [{\citenamefont {Beliaev}(1958)}]{Beliaev1958}%
  \BibitemOpen
  \bibfield  {author} {\bibinfo {author} {\bibfnamefont {S.~T.}\ \bibnamefont
  {Beliaev}},\ }\bibfield  {title} {\bibinfo {title} {Energy-{S}pectrum of a
  {N}on-ideal {B}ose gas},\ }\href@noop {} {\bibfield  {journal} {\bibinfo
  {journal} {Sov. Phys. JETP}\ }\textbf {\bibinfo {volume} {34}},\ \bibinfo
  {pages} {299} (\bibinfo {year} {1958})}\BibitemShut {NoStop}%
\bibitem [{\citenamefont {Hugenholtz}\ and\ \citenamefont
  {Pines}(1959)}]{Hugenholtz1959}%
  \BibitemOpen
  \bibfield  {author} {\bibinfo {author} {\bibfnamefont {N.~M.}\ \bibnamefont
  {Hugenholtz}}\ and\ \bibinfo {author} {\bibfnamefont {D.}~\bibnamefont
  {Pines}},\ }\bibfield  {title} {\bibinfo {title} {Ground-{S}tate {E}nergy and
  {E}xcitation {S}pectrum of a {S}ystem of {I}nteracting {B}osons},\ }\href
  {https://doi.org/10.1103/PhysRev.116.489} {\bibfield  {journal} {\bibinfo
  {journal} {Phys. Rev.}\ }\textbf {\bibinfo {volume} {116}},\ \bibinfo {pages}
  {489} (\bibinfo {year} {1959})}\BibitemShut {NoStop}%
\bibitem [{\citenamefont {Lieb}(1963{\natexlab{a}})}]{Lieb1963c}%
  \BibitemOpen
  \bibfield  {author} {\bibinfo {author} {\bibfnamefont {E.~H.}\ \bibnamefont
  {Lieb}},\ }\bibfield  {title} {\bibinfo {title} {Simplified {A}pproach to the
  {G}round-{S}tate {E}nergy of an {I}mperfect {B}ose {G}as},\ }\href
  {https://doi.org/10.1103/PhysRev.130.2518} {\bibfield  {journal} {\bibinfo
  {journal} {Phys. Rev.}\ }\textbf {\bibinfo {volume} {130}},\ \bibinfo {pages}
  {2518} (\bibinfo {year} {1963}{\natexlab{a}})}\BibitemShut {NoStop}%
\bibitem [{\citenamefont {Dalfovo}\ \emph {et~al.}(1999)\citenamefont
  {Dalfovo}, \citenamefont {Giorgini}, \citenamefont {Pitaevskii},\ and\
  \citenamefont {Stringari}}]{Dalfovo1999}%
  \BibitemOpen
  \bibfield  {author} {\bibinfo {author} {\bibfnamefont {F.}~\bibnamefont
  {Dalfovo}}, \bibinfo {author} {\bibfnamefont {S.}~\bibnamefont {Giorgini}},
  \bibinfo {author} {\bibfnamefont {L.~P.}\ \bibnamefont {Pitaevskii}},\ and\
  \bibinfo {author} {\bibfnamefont {S.}~\bibnamefont {Stringari}},\ }\bibfield
  {title} {\bibinfo {title} {Theory of {B}ose-{E}instein condensation in
  trapped gases},\ }\href {https://doi.org/10.1103/RevModPhys.71.463}
  {\bibfield  {journal} {\bibinfo  {journal} {Rev. Mod. Phys.}\ }\textbf
  {\bibinfo {volume} {71}},\ \bibinfo {pages} {463} (\bibinfo {year}
  {1999})}\BibitemShut {NoStop}%
\bibitem [{\citenamefont {Andersen}(2004)}]{Andersen2004}%
  \BibitemOpen
  \bibfield  {author} {\bibinfo {author} {\bibfnamefont {J.~O.}\ \bibnamefont
  {Andersen}},\ }\bibfield  {title} {\bibinfo {title} {Theory of the weakly
  interacting {B}ose gas},\ }\href {https://doi.org/10.1103/RevModPhys.76.599}
  {\bibfield  {journal} {\bibinfo  {journal} {Rev. Mod. Phys.}\ }\textbf
  {\bibinfo {volume} {76}},\ \bibinfo {pages} {599} (\bibinfo {year}
  {2004})}\BibitemShut {NoStop}%
\bibitem [{\citenamefont {Cazalilla}\ \emph {et~al.}(2011)\citenamefont
  {Cazalilla}, \citenamefont {Citro}, \citenamefont {Giamarchi}, \citenamefont
  {Orignac},\ and\ \citenamefont {Rigol}}]{Cazalilla2011}%
  \BibitemOpen
  \bibfield  {author} {\bibinfo {author} {\bibfnamefont {M.~A.}\ \bibnamefont
  {Cazalilla}}, \bibinfo {author} {\bibfnamefont {R.}~\bibnamefont {Citro}},
  \bibinfo {author} {\bibfnamefont {T.}~\bibnamefont {Giamarchi}}, \bibinfo
  {author} {\bibfnamefont {E.}~\bibnamefont {Orignac}},\ and\ \bibinfo {author}
  {\bibfnamefont {M.}~\bibnamefont {Rigol}},\ }\bibfield  {title} {\bibinfo
  {title} {One dimensional bosons: {F}rom condensed matter systems to ultracold
  gases},\ }\href {https://doi.org/10.1103/RevModPhys.83.1405} {\bibfield
  {journal} {\bibinfo  {journal} {Rev. Mod. Phys.}\ }\textbf {\bibinfo {volume}
  {83}},\ \bibinfo {pages} {1405} (\bibinfo {year} {2011})}\BibitemShut
  {NoStop}%
\bibitem [{\citenamefont {Lode}\ \emph {et~al.}(2020)\citenamefont {Lode},
  \citenamefont {L\'ev\^eque}, \citenamefont {Bojer~Madsen}, \citenamefont
  {Streltsov},\ and\ \citenamefont {Alon}}]{Lode2020}%
  \BibitemOpen
  \bibfield  {author} {\bibinfo {author} {\bibfnamefont {A.~U.~J.}\
  \bibnamefont {Lode}}, \bibinfo {author} {\bibfnamefont {C.}~\bibnamefont
  {L\'ev\^eque}}, \bibinfo {author} {\bibfnamefont {L.}~\bibnamefont
  {Bojer~Madsen}}, \bibinfo {author} {\bibfnamefont {A.~I.}\ \bibnamefont
  {Streltsov}},\ and\ \bibinfo {author} {\bibfnamefont {O.~E.}\ \bibnamefont
  {Alon}},\ }\bibfield  {title} {\bibinfo {title} {Multiconfigurational
  time-dependent {H}artree approaches for indistinguishable particles},\ }\href
  {https://doi.org/???} {\bibfield  {journal} {\bibinfo  {journal} {Rev. Mod.
  Phys.}\ }\textbf {\bibinfo {volume} {92}},\ \bibinfo {pages} {011001}
  (\bibinfo {year} {2020})}\BibitemShut {NoStop}%
\bibitem [{\citenamefont {Greiner}\ \emph {et~al.}(2003)\citenamefont
  {Greiner}, \citenamefont {Regal},\ and\ \citenamefont {Jin}}]{Greiner2003}%
  \BibitemOpen
  \bibfield  {author} {\bibinfo {author} {\bibfnamefont {M.}~\bibnamefont
  {Greiner}}, \bibinfo {author} {\bibfnamefont {C.~A.}\ \bibnamefont {Regal}},\
  and\ \bibinfo {author} {\bibfnamefont {D.~S.}\ \bibnamefont {Jin}},\
  }\bibfield  {title} {\bibinfo {title} {Emergence of a molecular
  {B}ose-{E}instein condensate from a {F}ermi gas},\ }\href
  {https://doi.org/10.1038/nature02199} {\bibfield  {journal} {\bibinfo
  {journal} {Nature}\ }\textbf {\bibinfo {volume} {426}},\ \bibinfo {pages}
  {537} (\bibinfo {year} {2003})}\BibitemShut {NoStop}%
\bibitem [{\citenamefont {Bartenstein}\ \emph {et~al.}(2004)\citenamefont
  {Bartenstein}, \citenamefont {Altmeyer}, \citenamefont {Riedl}, \citenamefont
  {Jochim}, \citenamefont {Chin}, \citenamefont {Hecker~Denschlag},\ and\
  \citenamefont {Grimm}}]{Bartenstein2004}%
  \BibitemOpen
  \bibfield  {author} {\bibinfo {author} {\bibfnamefont {M.}~\bibnamefont
  {Bartenstein}}, \bibinfo {author} {\bibfnamefont {A.}~\bibnamefont
  {Altmeyer}}, \bibinfo {author} {\bibfnamefont {S.}~\bibnamefont {Riedl}},
  \bibinfo {author} {\bibfnamefont {S.}~\bibnamefont {Jochim}}, \bibinfo
  {author} {\bibfnamefont {C.}~\bibnamefont {Chin}}, \bibinfo {author}
  {\bibfnamefont {J.}~\bibnamefont {Hecker~Denschlag}},\ and\ \bibinfo {author}
  {\bibfnamefont {R.}~\bibnamefont {Grimm}},\ }\bibfield  {title} {\bibinfo
  {title} {Crossover from a {M}olecular {B}ose-{E}instein {C}ondensate to a
  {D}egenerate {F}ermi {G}as},\ }\href
  {https://doi.org/10.1103/PhysRevLett.92.120401} {\bibfield  {journal}
  {\bibinfo  {journal} {Phys. Rev. Lett.}\ }\textbf {\bibinfo {volume} {92}},\
  \bibinfo {pages} {120401} (\bibinfo {year} {2004})}\BibitemShut {NoStop}%
\bibitem [{\citenamefont {Zwierlein}\ \emph {et~al.}(2004)\citenamefont
  {Zwierlein}, \citenamefont {Stan}, \citenamefont {Schunck}, \citenamefont
  {Raupach}, \citenamefont {Kerman},\ and\ \citenamefont
  {Ketterle}}]{Zwierlein2004}%
  \BibitemOpen
  \bibfield  {author} {\bibinfo {author} {\bibfnamefont {M.~W.}\ \bibnamefont
  {Zwierlein}}, \bibinfo {author} {\bibfnamefont {C.~A.}\ \bibnamefont {Stan}},
  \bibinfo {author} {\bibfnamefont {C.~H.}\ \bibnamefont {Schunck}}, \bibinfo
  {author} {\bibfnamefont {S.~M.~F.}\ \bibnamefont {Raupach}}, \bibinfo
  {author} {\bibfnamefont {A.~J.}\ \bibnamefont {Kerman}},\ and\ \bibinfo
  {author} {\bibfnamefont {W.}~\bibnamefont {Ketterle}},\ }\bibfield  {title}
  {\bibinfo {title} {Condensation of {P}airs of {F}ermionic {A}toms near a
  {F}eshbach {R}esonance},\ }\href
  {https://doi.org/10.1103/PhysRevLett.92.120403} {\bibfield  {journal}
  {\bibinfo  {journal} {Phys. Rev. Lett.}\ }\textbf {\bibinfo {volume} {92}},\
  \bibinfo {pages} {120403} (\bibinfo {year} {2004})}\BibitemShut {NoStop}%
\bibitem [{\citenamefont {Bourdel}\ \emph {et~al.}(2004)\citenamefont
  {Bourdel}, \citenamefont {Khaykovich}, \citenamefont {Cubizolles},
  \citenamefont {Zhang}, \citenamefont {Chevy}, \citenamefont {Teichmann},
  \citenamefont {Tarruell}, \citenamefont {Kokkelmans},\ and\ \citenamefont
  {Salomon}}]{Bourdel2004}%
  \BibitemOpen
  \bibfield  {author} {\bibinfo {author} {\bibfnamefont {T.}~\bibnamefont
  {Bourdel}}, \bibinfo {author} {\bibfnamefont {L.}~\bibnamefont {Khaykovich}},
  \bibinfo {author} {\bibfnamefont {J.}~\bibnamefont {Cubizolles}}, \bibinfo
  {author} {\bibfnamefont {J.}~\bibnamefont {Zhang}}, \bibinfo {author}
  {\bibfnamefont {F.}~\bibnamefont {Chevy}}, \bibinfo {author} {\bibfnamefont
  {M.}~\bibnamefont {Teichmann}}, \bibinfo {author} {\bibfnamefont
  {L.}~\bibnamefont {Tarruell}}, \bibinfo {author} {\bibfnamefont {S.~J. J.
  M.~F.}\ \bibnamefont {Kokkelmans}},\ and\ \bibinfo {author} {\bibfnamefont
  {C.}~\bibnamefont {Salomon}},\ }\bibfield  {title} {\bibinfo {title}
  {Experimental {S}tudy of the {B}{E}{C}-{B}{C}{S} {C}rossover {R}egion in
  {L}ithium 6},\ }\href {https://doi.org/10.1103/PhysRevLett.93.050401}
  {\bibfield  {journal} {\bibinfo  {journal} {Phys. Rev. Lett.}\ }\textbf
  {\bibinfo {volume} {93}},\ \bibinfo {pages} {050401} (\bibinfo {year}
  {2004})}\BibitemShut {NoStop}%
\bibitem [{\citenamefont {Greiner}\ \emph {et~al.}(2001)\citenamefont
  {Greiner}, \citenamefont {Bloch}, \citenamefont {Mandel}, \citenamefont
  {H\"ansch},\ and\ \citenamefont {Esslinger}}]{Greiner2001}%
  \BibitemOpen
  \bibfield  {author} {\bibinfo {author} {\bibfnamefont {M.}~\bibnamefont
  {Greiner}}, \bibinfo {author} {\bibfnamefont {I.}~\bibnamefont {Bloch}},
  \bibinfo {author} {\bibfnamefont {O.}~\bibnamefont {Mandel}}, \bibinfo
  {author} {\bibfnamefont {T.~W.}\ \bibnamefont {H\"ansch}},\ and\ \bibinfo
  {author} {\bibfnamefont {T.}~\bibnamefont {Esslinger}},\ }\bibfield  {title}
  {\bibinfo {title} {Exploring {P}hase {C}oherence in a 2{D} {L}attice of
  {B}ose-{E}instein {C}ondensates},\ }\href
  {https://doi.org/10.1103/PhysRevLett.87.160405} {\bibfield  {journal}
  {\bibinfo  {journal} {Phys. Rev. Lett.}\ }\textbf {\bibinfo {volume} {87}},\
  \bibinfo {pages} {160405} (\bibinfo {year} {2001})}\BibitemShut {NoStop}%
\bibitem [{\citenamefont {Dettmer}\ \emph {et~al.}(2001)\citenamefont
  {Dettmer}, \citenamefont {Hellweg}, \citenamefont {Ryytty}, \citenamefont
  {Arlt}, \citenamefont {Ertmer}, \citenamefont {Sengstock}, \citenamefont
  {Petrov}, \citenamefont {Shlyapnikov}, \citenamefont {Kreutzmann},
  \citenamefont {Santos},\ and\ \citenamefont {Lewenstein}}]{Dettmer2001}%
  \BibitemOpen
  \bibfield  {author} {\bibinfo {author} {\bibfnamefont {S.}~\bibnamefont
  {Dettmer}}, \bibinfo {author} {\bibfnamefont {D.}~\bibnamefont {Hellweg}},
  \bibinfo {author} {\bibfnamefont {P.}~\bibnamefont {Ryytty}}, \bibinfo
  {author} {\bibfnamefont {J.~J.}\ \bibnamefont {Arlt}}, \bibinfo {author}
  {\bibfnamefont {W.}~\bibnamefont {Ertmer}}, \bibinfo {author} {\bibfnamefont
  {K.}~\bibnamefont {Sengstock}}, \bibinfo {author} {\bibfnamefont {D.~S.}\
  \bibnamefont {Petrov}}, \bibinfo {author} {\bibfnamefont {G.~V.}\
  \bibnamefont {Shlyapnikov}}, \bibinfo {author} {\bibfnamefont
  {H.}~\bibnamefont {Kreutzmann}}, \bibinfo {author} {\bibfnamefont
  {L.}~\bibnamefont {Santos}},\ and\ \bibinfo {author} {\bibfnamefont
  {M.}~\bibnamefont {Lewenstein}},\ }\bibfield  {title} {\bibinfo {title}
  {Observation of {P}hase {F}luctuations in {E}longated {B}ose-{E}instein
  {C}ondensates},\ }\href {https://doi.org/10.1103/PhysRevLett.87.160406}
  {\bibfield  {journal} {\bibinfo  {journal} {Phys. Rev. Lett.}\ }\textbf
  {\bibinfo {volume} {87}},\ \bibinfo {pages} {160406} (\bibinfo {year}
  {2001})}\BibitemShut {NoStop}%
\bibitem [{\citenamefont {G\"orlitz}\ \emph {et~al.}(2001)\citenamefont
  {G\"orlitz}, \citenamefont {Vogels}, \citenamefont {Leanhardt}, \citenamefont
  {Raman}, \citenamefont {Gustavson}, \citenamefont {Abo-Shaeer}, \citenamefont
  {Chikkatur}, \citenamefont {Gupta}, \citenamefont {Inouye}, \citenamefont
  {Rosenband},\ and\ \citenamefont {Ketterle}}]{Goerlitz2001}%
  \BibitemOpen
  \bibfield  {author} {\bibinfo {author} {\bibfnamefont {A.}~\bibnamefont
  {G\"orlitz}}, \bibinfo {author} {\bibfnamefont {J.~M.}\ \bibnamefont
  {Vogels}}, \bibinfo {author} {\bibfnamefont {A.~E.}\ \bibnamefont
  {Leanhardt}}, \bibinfo {author} {\bibfnamefont {C.}~\bibnamefont {Raman}},
  \bibinfo {author} {\bibfnamefont {T.~L.}\ \bibnamefont {Gustavson}}, \bibinfo
  {author} {\bibfnamefont {J.~R.}\ \bibnamefont {Abo-Shaeer}}, \bibinfo
  {author} {\bibfnamefont {A.~P.}\ \bibnamefont {Chikkatur}}, \bibinfo {author}
  {\bibfnamefont {S.}~\bibnamefont {Gupta}}, \bibinfo {author} {\bibfnamefont
  {S.}~\bibnamefont {Inouye}}, \bibinfo {author} {\bibfnamefont
  {T.}~\bibnamefont {Rosenband}},\ and\ \bibinfo {author} {\bibfnamefont
  {W.}~\bibnamefont {Ketterle}},\ }\bibfield  {title} {\bibinfo {title}
  {Realization of {B}ose-{E}instein {C}ondensates in {L}ower {D}imensions},\
  }\href {https://doi.org/10.1103/PhysRevLett.87.130402} {\bibfield  {journal}
  {\bibinfo  {journal} {Phys. Rev. Lett.}\ }\textbf {\bibinfo {volume} {87}},\
  \bibinfo {pages} {130402} (\bibinfo {year} {2001})}\BibitemShut {NoStop}%
\bibitem [{\citenamefont {Orzel}\ \emph {et~al.}(2001)\citenamefont {Orzel},
  \citenamefont {Tuchman}, \citenamefont {Fenselau}, \citenamefont {Yasuda},\
  and\ \citenamefont {Kasevich}}]{Orzel2001}%
  \BibitemOpen
  \bibfield  {author} {\bibinfo {author} {\bibfnamefont {C.}~\bibnamefont
  {Orzel}}, \bibinfo {author} {\bibfnamefont {A.~K.}\ \bibnamefont {Tuchman}},
  \bibinfo {author} {\bibfnamefont {M.~L.}\ \bibnamefont {Fenselau}}, \bibinfo
  {author} {\bibfnamefont {M.}~\bibnamefont {Yasuda}},\ and\ \bibinfo {author}
  {\bibfnamefont {M.~A.}\ \bibnamefont {Kasevich}},\ }\bibfield  {title}
  {\bibinfo {title} {Squeezed {S}tates in a {B}ose-{E}instein {C}ondensate},\
  }\href {https://doi.org/10.1126/science.1058149} {\bibfield  {journal}
  {\bibinfo  {journal} {Science}\ }\textbf {\bibinfo {volume} {291}},\ \bibinfo
  {pages} {2386} (\bibinfo {year} {2001})}\BibitemShut {NoStop}%
\bibitem [{\citenamefont {Paredes}\ \emph {et~al.}(2004)\citenamefont
  {Paredes}, \citenamefont {Widera}, \citenamefont {Murg}, \citenamefont
  {Mandel}, \citenamefont {F\"olling}, \citenamefont {Cirac}, \citenamefont
  {Shlyapnikov}, \citenamefont {H\"ansch},\ and\ \citenamefont
  {Bloch}}]{Paredes2004}%
  \BibitemOpen
  \bibfield  {author} {\bibinfo {author} {\bibfnamefont {B.}~\bibnamefont
  {Paredes}}, \bibinfo {author} {\bibfnamefont {A.}~\bibnamefont {Widera}},
  \bibinfo {author} {\bibfnamefont {V.}~\bibnamefont {Murg}}, \bibinfo {author}
  {\bibfnamefont {O.}~\bibnamefont {Mandel}}, \bibinfo {author} {\bibfnamefont
  {S.}~\bibnamefont {F\"olling}}, \bibinfo {author} {\bibfnamefont
  {I.}~\bibnamefont {Cirac}}, \bibinfo {author} {\bibfnamefont {G.~V.}\
  \bibnamefont {Shlyapnikov}}, \bibinfo {author} {\bibfnamefont {T.~W.}\
  \bibnamefont {H\"ansch}},\ and\ \bibinfo {author} {\bibfnamefont
  {I.}~\bibnamefont {Bloch}},\ }\bibfield  {title} {\bibinfo {title}
  {Tonks-{G}irardeau gas of ultracold atoms in an optical lattice},\ }\href
  {https://doi.org/10.1038/nature02530} {\bibfield  {journal} {\bibinfo
  {journal} {Nature}\ }\textbf {\bibinfo {volume} {429}},\ \bibinfo {pages}
  {277} (\bibinfo {year} {2004})}\BibitemShut {NoStop}%
\bibitem [{\citenamefont {Kinoshita}\ \emph {et~al.}(2004)\citenamefont
  {Kinoshita}, \citenamefont {Wenger},\ and\ \citenamefont
  {Weiss}}]{Kinoshita2004}%
  \BibitemOpen
  \bibfield  {author} {\bibinfo {author} {\bibfnamefont {T.}~\bibnamefont
  {Kinoshita}}, \bibinfo {author} {\bibfnamefont {T.}~\bibnamefont {Wenger}},\
  and\ \bibinfo {author} {\bibfnamefont {D.~S.}\ \bibnamefont {Weiss}},\
  }\bibfield  {title} {\bibinfo {title} {Observation of a {O}ne-{D}imensional
  {T}onks-{G}irardeau {G}as},\ }\href {https://doi.org/10.1126/science.1100700}
  {\bibfield  {journal} {\bibinfo  {journal} {Science}\ }\textbf {\bibinfo
  {volume} {305}},\ \bibinfo {pages} {1125} (\bibinfo {year}
  {2004})}\BibitemShut {NoStop}%
\bibitem [{\citenamefont {St\"oferle}\ \emph {et~al.}(2004)\citenamefont
  {St\"oferle}, \citenamefont {Moritz}, \citenamefont {Schori}, \citenamefont
  {K\"ohl},\ and\ \citenamefont {Esslinger}}]{Stoeferle2004}%
  \BibitemOpen
  \bibfield  {author} {\bibinfo {author} {\bibfnamefont {T.}~\bibnamefont
  {St\"oferle}}, \bibinfo {author} {\bibfnamefont {H.}~\bibnamefont {Moritz}},
  \bibinfo {author} {\bibfnamefont {C.}~\bibnamefont {Schori}}, \bibinfo
  {author} {\bibfnamefont {M.}~\bibnamefont {K\"ohl}},\ and\ \bibinfo {author}
  {\bibfnamefont {T.}~\bibnamefont {Esslinger}},\ }\bibfield  {title} {\bibinfo
  {title} {Transition from a {S}trongly {I}nteracting 1{D} {S}uperfluid to a
  {M}ott insulator},\ }\href {https://doi.org/10.1103/PhysRevLett.92.130403}
  {\bibfield  {journal} {\bibinfo  {journal} {Phys. Rev. Lett.}\ }\textbf
  {\bibinfo {volume} {92}},\ \bibinfo {pages} {130403} (\bibinfo {year}
  {2004})}\BibitemShut {NoStop}%
\bibitem [{\citenamefont {Lenard}(1964)}]{Lenard1964}%
  \BibitemOpen
  \bibfield  {author} {\bibinfo {author} {\bibfnamefont {A.}~\bibnamefont
  {Lenard}},\ }\bibfield  {title} {\bibinfo {title} {Momentum distribution in
  the ground state of the one-dimensional systems of impenetrable bosons},\
  }\href {https://doi.org/10.1063/1.1704196} {\bibfield  {journal} {\bibinfo
  {journal} {J. Math. Phys.}\ }\textbf {\bibinfo {volume} {5}},\ \bibinfo
  {pages} {930} (\bibinfo {year} {1964})}\BibitemShut {NoStop}%
\bibitem [{\citenamefont {Popov}(1972)}]{Popov1972}%
  \BibitemOpen
  \bibfield  {author} {\bibinfo {author} {\bibfnamefont {V.~N.}\ \bibnamefont
  {Popov}},\ }\bibfield  {title} {\bibinfo {title} {On the theory of the
  superfluidity of two- and one-dimensional bose systems},\ }\href
  {https://doi.org/10.1007/BF01028373} {\bibfield  {journal} {\bibinfo
  {journal} {Theor. Math. Phys.}\ }\textbf {\bibinfo {volume} {11}},\ \bibinfo
  {pages} {565} (\bibinfo {year} {1972})}\BibitemShut {NoStop}%
\bibitem [{\citenamefont {Widom}(1973)}]{Widom1973}%
  \BibitemOpen
  \bibfield  {author} {\bibinfo {author} {\bibfnamefont {H.}~\bibnamefont
  {Widom}},\ }\bibfield  {title} {\bibinfo {title} {Toeplitz {D}eterminants
  with {S}ingular {G}enerating {F}unctions},\ }\href
  {https://doi.org/10.2307/2373789} {\bibfield  {journal} {\bibinfo  {journal}
  {Am. J. Math.}\ }\textbf {\bibinfo {volume} {95}},\ \bibinfo {pages} {333}
  (\bibinfo {year} {1973})}\BibitemShut {NoStop}%
\bibitem [{\citenamefont {Schwartz}(1977)}]{Schwartz1977}%
  \BibitemOpen
  \bibfield  {author} {\bibinfo {author} {\bibfnamefont {M.}~\bibnamefont
  {Schwartz}},\ }\bibfield  {title} {\bibinfo {title} {Off-diagonal long-range
  behavior of interacting {B}ose systems},\ }\href
  {https://doi.org/10.1103/PhysRevB.15.1399} {\bibfield  {journal} {\bibinfo
  {journal} {Phys. Rev. B}\ }\textbf {\bibinfo {volume} {15}},\ \bibinfo
  {pages} {1399} (\bibinfo {year} {1977})}\BibitemShut {NoStop}%
\bibitem [{\citenamefont {Girardeau}\ and\ \citenamefont
  {Wright}(2001)}]{Girardeau2001}%
  \BibitemOpen
  \bibfield  {author} {\bibinfo {author} {\bibfnamefont {M.~D.}\ \bibnamefont
  {Girardeau}}\ and\ \bibinfo {author} {\bibfnamefont {E.~M.}\ \bibnamefont
  {Wright}},\ }\bibfield  {title} {\bibinfo {title} {Bose-{F}ermi {V}ariational
  {T}heory of the {B}ose-{E}instein {C}ondensate {C}rossover to the {T}onks
  {G}as},\ }\href {https://doi.org/10.1103/PhysRevLett.87.210401} {\bibfield
  {journal} {\bibinfo  {journal} {Phys. Rev. Lett.}\ }\textbf {\bibinfo
  {volume} {87}},\ \bibinfo {pages} {210401} (\bibinfo {year}
  {2001})}\BibitemShut {NoStop}%
\bibitem [{\citenamefont {Forrester}\ \emph {et~al.}(2003)\citenamefont
  {Forrester}, \citenamefont {Frankel}, \citenamefont {Garoni},\ and\
  \citenamefont {Witte}}]{Forrester2003b}%
  \BibitemOpen
  \bibfield  {author} {\bibinfo {author} {\bibfnamefont {P.~J.}\ \bibnamefont
  {Forrester}}, \bibinfo {author} {\bibfnamefont {N.~E.}\ \bibnamefont
  {Frankel}}, \bibinfo {author} {\bibfnamefont {T.~M.}\ \bibnamefont
  {Garoni}},\ and\ \bibinfo {author} {\bibfnamefont {N.~S.}\ \bibnamefont
  {Witte}},\ }\bibfield  {title} {\bibinfo {title} {Finite one-dimensional
  impenetrable bose systems: Occupation numbers},\ }\href
  {https://doi.org/10.1103/PhysRevA.67.043607} {\bibfield  {journal} {\bibinfo
  {journal} {Phys. Rev A}\ }\textbf {\bibinfo {volume} {67}},\ \bibinfo {pages}
  {043607} (\bibinfo {year} {2003})}\BibitemShut {NoStop}%
\bibitem [{\citenamefont {Gangardt}(2004)}]{Gangardt2004}%
  \BibitemOpen
  \bibfield  {author} {\bibinfo {author} {\bibfnamefont {D.~M.}\ \bibnamefont
  {Gangardt}},\ }\bibfield  {title} {\bibinfo {title} {Universal correlations
  of trapped one-dimensional impenetrable bosons},\ }\href
  {https://doi.org/10.1088/0305-4470/37/40/002} {\bibfield  {journal} {\bibinfo
   {journal} {J. Phys. A: Math. Gen.}\ }\textbf {\bibinfo {volume} {37}},\
  \bibinfo {pages} {9335} (\bibinfo {year} {2004})}\BibitemShut {NoStop}%
\bibitem [{\citenamefont {Rigol}\ and\ \citenamefont
  {Muramatsu}(2004{\natexlab{a}})}]{RigolHCB1d2004}%
  \BibitemOpen
  \bibfield  {author} {\bibinfo {author} {\bibfnamefont {M.}~\bibnamefont
  {Rigol}}\ and\ \bibinfo {author} {\bibfnamefont {A.}~\bibnamefont
  {Muramatsu}},\ }\bibfield  {title} {\bibinfo {title} {Universal properties of
  hard-core bosons confined on one-dimensional lattices},\ }\href
  {https://doi.org/10.1103/PhysRevA.70.031603} {\bibfield  {journal} {\bibinfo
  {journal} {Phys. Rev. A}\ }\textbf {\bibinfo {volume} {70}},\ \bibinfo
  {pages} {031603R} (\bibinfo {year} {2004}{\natexlab{a}})}\BibitemShut
  {NoStop}%
\bibitem [{\citenamefont {Rigol}\ and\ \citenamefont
  {Muramatsu}(2004{\natexlab{b}})}]{RigolHCB1dPRL}%
  \BibitemOpen
  \bibfield  {author} {\bibinfo {author} {\bibfnamefont {M.}~\bibnamefont
  {Rigol}}\ and\ \bibinfo {author} {\bibfnamefont {A.}~\bibnamefont
  {Muramatsu}},\ }\bibfield  {title} {\bibinfo {title} {Emergence of
  quasicondensates of hard-core bosons at finite momentum},\ }\href
  {https://doi.org/10.1103/PhysRevLett.93.230404} {\bibfield  {journal}
  {\bibinfo  {journal} {Phys. Rev. Lett.}\ }\textbf {\bibinfo {volume} {93}},\
  \bibinfo {pages} {230404} (\bibinfo {year} {2004}{\natexlab{b}})}\BibitemShut
  {NoStop}%
\bibitem [{\citenamefont {Lieb}\ and\ \citenamefont
  {Liniger}(1963)}]{Lieb1963a}%
  \BibitemOpen
  \bibfield  {author} {\bibinfo {author} {\bibfnamefont {E.~H.}\ \bibnamefont
  {Lieb}}\ and\ \bibinfo {author} {\bibfnamefont {W.}~\bibnamefont {Liniger}},\
  }\bibfield  {title} {\bibinfo {title} {Exact {A}nalysis of an {I}nteracting
  {B}ose {G}as. {I}. {T}he {G}eneral {S}olution and the {G}round {S}tate},\
  }\href {https://doi.org/10.1103/PhysRev.130.1605} {\bibfield  {journal}
  {\bibinfo  {journal} {Phys. Rev.}\ }\textbf {\bibinfo {volume} {130}},\
  \bibinfo {pages} {1605} (\bibinfo {year} {1963})}\BibitemShut {NoStop}%
\bibitem [{\citenamefont {Lieb}(1963{\natexlab{b}})}]{Lieb1963b}%
  \BibitemOpen
  \bibfield  {author} {\bibinfo {author} {\bibfnamefont {E.~H.}\ \bibnamefont
  {Lieb}},\ }\bibfield  {title} {\bibinfo {title} {Exact {A}nalysis of an
  {I}nteracting {B}ose {G}as. {II}. {T}he {E}xcitation {S}pectrum},\ }\href
  {https://doi.org/10.1103/PhysRev.130.1616} {\bibfield  {journal} {\bibinfo
  {journal} {Phys. Rev.}\ }\textbf {\bibinfo {volume} {130}},\ \bibinfo {pages}
  {1616} (\bibinfo {year} {1963}{\natexlab{b}})}\BibitemShut {NoStop}%
\bibitem [{\citenamefont {Girardeau}(1960)}]{Girardeau1960}%
  \BibitemOpen
  \bibfield  {author} {\bibinfo {author} {\bibfnamefont {M.}~\bibnamefont
  {Girardeau}},\ }\bibfield  {title} {\bibinfo {title} {Relationship between
  systems of impenetrable bosons and fermions in one dimension},\ }\href
  {https://doi.org/10.1063/1.1703687} {\bibfield  {journal} {\bibinfo
  {journal} {J. Math. Phys.}\ }\textbf {\bibinfo {volume} {1}},\ \bibinfo
  {pages} {516} (\bibinfo {year} {1960})}\BibitemShut {NoStop}%
\bibitem [{\citenamefont {Sachdev}(1999)}]{Sachdev1999}%
  \BibitemOpen
  \bibfield  {author} {\bibinfo {author} {\bibfnamefont {S.}~\bibnamefont
  {Sachdev}},\ }\href@noop {} {\emph {\bibinfo {title} {Quantum Phase
  Transitions}}}\ (\bibinfo  {publisher} {Cambridge University Press,
  Cambridge},\ \bibinfo {year} {1999})\BibitemShut {NoStop}%
\bibitem [{\citenamefont {Osterloh}\ \emph {et~al.}(2002)\citenamefont
  {Osterloh}, \citenamefont {Amico},\ and\ \citenamefont
  {Fazio}}]{Osterloh2002}%
  \BibitemOpen
  \bibfield  {author} {\bibinfo {author} {\bibfnamefont {A.}~\bibnamefont
  {Osterloh}}, \bibinfo {author} {\bibfnamefont {L.}~\bibnamefont {Amico}},\
  and\ \bibinfo {author} {\bibfnamefont {R.}~\bibnamefont {Fazio}},\ }\bibfield
   {title} {\bibinfo {title} {Scaling of entanglement close to a quantum phase
  transition},\ }\href {https://doi.org/10.1038/416608a} {\bibfield  {journal}
  {\bibinfo  {journal} {Nature}\ }\textbf {\bibinfo {volume} {416}},\ \bibinfo
  {pages} {608} (\bibinfo {year} {2002})}\BibitemShut {NoStop}%
\bibitem [{\citenamefont {Osborne}\ and\ \citenamefont
  {Nielsen}(2019)}]{Osborne2002}%
  \BibitemOpen
  \bibfield  {author} {\bibinfo {author} {\bibfnamefont {T.~J.}\ \bibnamefont
  {Osborne}}\ and\ \bibinfo {author} {\bibfnamefont {M.~A.}\ \bibnamefont
  {Nielsen}},\ }\bibfield  {title} {\bibinfo {title} {Entanglement in a simple
  quantum phase transition},\ }\href {https://doi.org/???} {\bibfield
  {journal} {\bibinfo  {journal} {Phys. Rev. A}\ }\textbf {\bibinfo {volume}
  {66}},\ \bibinfo {pages} {032110} (\bibinfo {year} {2019})}\BibitemShut
  {NoStop}%
\bibitem [{\citenamefont {Das}\ \emph {et~al.}(2002)\citenamefont {Das},
  \citenamefont {Girardeau},\ and\ \citenamefont {Wright}}]{Das2002}%
  \BibitemOpen
  \bibfield  {author} {\bibinfo {author} {\bibfnamefont {K.~K.}\ \bibnamefont
  {Das}}, \bibinfo {author} {\bibfnamefont {M.~D.}\ \bibnamefont {Girardeau}},\
  and\ \bibinfo {author} {\bibfnamefont {E.~M.}\ \bibnamefont {Wright}},\
  }\bibfield  {title} {\bibinfo {title} {Crossover from {O}ne to {T}hree
  {D}imensions for a {G}as of {H}ard-{C}ore {B}osons},\ }\href
  {https://doi.org/10.1103/PhysRevLett.89.110402} {\bibfield  {journal}
  {\bibinfo  {journal} {Phys. Rev. Lett.}\ }\textbf {\bibinfo {volume} {89}},\
  \bibinfo {pages} {110402} (\bibinfo {year} {2002})}\BibitemShut {NoStop}%
\bibitem [{\citenamefont {Salasnich}\ \emph {et~al.}(2004)\citenamefont
  {Salasnich}, \citenamefont {Parola},\ and\ \citenamefont
  {Reatto}}]{Salasnich2004}%
  \BibitemOpen
  \bibfield  {author} {\bibinfo {author} {\bibfnamefont {L.}~\bibnamefont
  {Salasnich}}, \bibinfo {author} {\bibfnamefont {A.}~\bibnamefont {Parola}},\
  and\ \bibinfo {author} {\bibfnamefont {L.}~\bibnamefont {Reatto}},\
  }\bibfield  {title} {\bibinfo {title} {Transition from three dimensions to
  one dimension in {B}ose gases at zero temperature},\ }\href
  {https://doi.org/10.1103/PhysRevA.70.013606} {\bibfield  {journal} {\bibinfo
  {journal} {Phys. Rev. A}\ }\textbf {\bibinfo {volume} {70}},\ \bibinfo
  {pages} {013606} (\bibinfo {year} {2004})}\BibitemShut {NoStop}%
\bibitem [{\citenamefont {Salasnich}\ \emph {et~al.}(2005)\citenamefont
  {Salasnich}, \citenamefont {Parola},\ and\ \citenamefont
  {Reatto}}]{Salasnich2005}%
  \BibitemOpen
  \bibfield  {author} {\bibinfo {author} {\bibfnamefont {L.}~\bibnamefont
  {Salasnich}}, \bibinfo {author} {\bibfnamefont {A.}~\bibnamefont {Parola}},\
  and\ \bibinfo {author} {\bibfnamefont {L.}~\bibnamefont {Reatto}},\
  }\bibfield  {title} {\bibinfo {title} {Quasi-one-dimensional bosons in
  three-dimensional traps: {F}rom strong-coupling to weak-coupling regime},\
  }\href {https://doi.org/10.1103/PhysRevA.72.025602} {\bibfield  {journal}
  {\bibinfo  {journal} {Phys. Rev. A}\ }\textbf {\bibinfo {volume} {72}},\
  \bibinfo {pages} {025602} (\bibinfo {year} {2005})}\BibitemShut {NoStop}%
\bibitem [{\citenamefont {Petrov}\ \emph {et~al.}(2000)\citenamefont {Petrov},
  \citenamefont {Shlyapnikov},\ and\ \citenamefont {Walraven}}]{Petrov2000}%
  \BibitemOpen
  \bibfield  {author} {\bibinfo {author} {\bibfnamefont {D.~S.}\ \bibnamefont
  {Petrov}}, \bibinfo {author} {\bibfnamefont {G.~V.}\ \bibnamefont
  {Shlyapnikov}},\ and\ \bibinfo {author} {\bibfnamefont {J.~T.~M.}\
  \bibnamefont {Walraven}},\ }\bibfield  {title} {\bibinfo {title} {Regimes of
  {Q}uantum {D}egeneracy in {T}rapped 1{D} {G}ases},\ }\href
  {https://doi.org/10.1103/PhysRevLett.85.3745} {\bibfield  {journal} {\bibinfo
   {journal} {Phys. Rev. Lett.}\ }\textbf {\bibinfo {volume} {85}},\ \bibinfo
  {pages} {3745} (\bibinfo {year} {2000})}\BibitemShut {NoStop}%
\bibitem [{\citenamefont {Fisher}\ \emph {et~al.}(1989)\citenamefont {Fisher},
  \citenamefont {Weichman}, \citenamefont {Grinstein},\ and\ \citenamefont
  {Fisher}}]{Fisher1989}%
  \BibitemOpen
  \bibfield  {author} {\bibinfo {author} {\bibfnamefont {M.~P.~A.}\
  \bibnamefont {Fisher}}, \bibinfo {author} {\bibfnamefont {P.~B.}\
  \bibnamefont {Weichman}}, \bibinfo {author} {\bibfnamefont {G.}~\bibnamefont
  {Grinstein}},\ and\ \bibinfo {author} {\bibfnamefont {D.~S.}\ \bibnamefont
  {Fisher}},\ }\bibfield  {title} {\bibinfo {title} {Boson localization and the
  superfluid-insulator transition},\ }\href
  {https://doi.org/10.1103/PhysRevB.40.546} {\bibfield  {journal} {\bibinfo
  {journal} {Phys. Rev. B}\ }\textbf {\bibinfo {volume} {40}},\ \bibinfo
  {pages} {546} (\bibinfo {year} {1989})}\BibitemShut {NoStop}%
\bibitem [{\citenamefont {Jaksch}\ \emph {et~al.}(1998)\citenamefont {Jaksch},
  \citenamefont {Bruder}, \citenamefont {Cirac}, \citenamefont {Gardiner},\
  and\ \citenamefont {Zoller}}]{Jaksch1998}%
  \BibitemOpen
  \bibfield  {author} {\bibinfo {author} {\bibfnamefont {D.}~\bibnamefont
  {Jaksch}}, \bibinfo {author} {\bibfnamefont {C.}~\bibnamefont {Bruder}},
  \bibinfo {author} {\bibfnamefont {J.~I.}\ \bibnamefont {Cirac}}, \bibinfo
  {author} {\bibfnamefont {C.~W.}\ \bibnamefont {Gardiner}},\ and\ \bibinfo
  {author} {\bibfnamefont {P.}~\bibnamefont {Zoller}},\ }\bibfield  {title}
  {\bibinfo {title} {Cold {B}osonic {A}toms in {O}ptical {L}attices},\ }\href
  {https://doi.org/???} {\bibfield  {journal} {\bibinfo  {journal} {Phys. Rev.
  Lett.}\ }\textbf {\bibinfo {volume} {81}},\ \bibinfo {pages} {3108} (\bibinfo
  {year} {1998})}\BibitemShut {NoStop}%
\bibitem [{\citenamefont {Matsubara}\ and\ \citenamefont
  {Matsuda}(1956)}]{Matsubara1956}%
  \BibitemOpen
  \bibfield  {author} {\bibinfo {author} {\bibfnamefont {T.}~\bibnamefont
  {Matsubara}}\ and\ \bibinfo {author} {\bibfnamefont {H.}~\bibnamefont
  {Matsuda}},\ }\bibfield  {title} {\bibinfo {title} {A lattice model of liquid
  {H}elium, {I}},\ }\href {https://doi.org/10.1143/PTP.16.569} {\bibfield
  {journal} {\bibinfo  {journal} {Prog. Theor. Phys.}\ }\textbf {\bibinfo
  {volume} {16}},\ \bibinfo {pages} {569} (\bibinfo {year} {1956})}\BibitemShut
  {NoStop}%
\bibitem [{\citenamefont {Matsuda}\ and\ \citenamefont
  {Matsubara}(1957)}]{Matsuda1957}%
  \BibitemOpen
  \bibfield  {author} {\bibinfo {author} {\bibfnamefont {H.}~\bibnamefont
  {Matsuda}}\ and\ \bibinfo {author} {\bibfnamefont {T.}~\bibnamefont
  {Matsubara}},\ }\bibfield  {title} {\bibinfo {title} {A lattice model of
  liquid {H}elium, {II}},\ }\href {https://doi.org/10.1143/PTP.17.19}
  {\bibfield  {journal} {\bibinfo  {journal} {Prog. Theor. Phys.}\ }\textbf
  {\bibinfo {volume} {17}},\ \bibinfo {pages} {19} (\bibinfo {year}
  {1957})}\BibitemShut {NoStop}%
\bibitem [{\citenamefont {T\'oth}(1990)}]{Toth1990}%
  \BibitemOpen
  \bibfield  {author} {\bibinfo {author} {\bibfnamefont {B.}~\bibnamefont
  {T\'oth}},\ }\bibfield  {title} {\bibinfo {title} {Phase transitions in an
  interacting bose system. an application of the theory of {V}entsel' and
  {F}reidlin},\ }\href {https://doi.org/10.1007/BF01027300} {\bibfield
  {journal} {\bibinfo  {journal} {J. Stat. Phys.}\ }\textbf {\bibinfo {volume}
  {61}},\ \bibinfo {pages} {749} (\bibinfo {year} {1990})}\BibitemShut
  {NoStop}%
\bibitem [{\citenamefont {Penrose}(1991)}]{Penrose1991}%
  \BibitemOpen
  \bibfield  {author} {\bibinfo {author} {\bibfnamefont {O.}~\bibnamefont
  {Penrose}},\ }\bibfield  {title} {\bibinfo {title} {Bose-{E}instein
  condensation in an exactly soluble system of interacting particles},\ }\href
  {https://doi.org/10.1007/BF01029210} {\bibfield  {journal} {\bibinfo
  {journal} {J. Stat. Phys.}\ }\textbf {\bibinfo {volume} {63}},\ \bibinfo
  {pages} {761} (\bibinfo {year} {1991})}\BibitemShut {NoStop}%
\bibitem [{\citenamefont {Kirson}(2000)}]{Kirson2000}%
  \BibitemOpen
  \bibfield  {author} {\bibinfo {author} {\bibfnamefont {M.~W.}\ \bibnamefont
  {Kirson}},\ }\bibfield  {title} {\bibinfo {title} {Bose-{E}instein
  condensation in an exactly solvable model for strongly interacting bosons},\
  }\href {https://doi.org/10.1088/0305-4470/33/4/307} {\bibfield  {journal}
  {\bibinfo  {journal} {J. Phys. A: Math. Gen.}\ }\textbf {\bibinfo {volume}
  {33}},\ \bibinfo {pages} {731} (\bibinfo {year} {2000})}\BibitemShut
  {NoStop}%
\bibitem [{\citenamefont {G\"unter}\ \emph {et~al.}(2013)\citenamefont
  {G\"unter}, \citenamefont {Schempp}, \citenamefont {Robert-de Saint-Vincent},
  \citenamefont {Gavryusev}, \citenamefont {Helmrich}, \citenamefont {Hofmann},
  \citenamefont {Whitlock},\ and\ \citenamefont {Weidem\"uller}}]{Guenter2013}%
  \BibitemOpen
  \bibfield  {author} {\bibinfo {author} {\bibfnamefont {G.}~\bibnamefont
  {G\"unter}}, \bibinfo {author} {\bibfnamefont {H.}~\bibnamefont {Schempp}},
  \bibinfo {author} {\bibfnamefont {M.}~\bibnamefont {Robert-de
  Saint-Vincent}}, \bibinfo {author} {\bibfnamefont {V.}~\bibnamefont
  {Gavryusev}}, \bibinfo {author} {\bibfnamefont {S.}~\bibnamefont {Helmrich}},
  \bibinfo {author} {\bibfnamefont {C.~S.}\ \bibnamefont {Hofmann}}, \bibinfo
  {author} {\bibfnamefont {S.}~\bibnamefont {Whitlock}},\ and\ \bibinfo
  {author} {\bibfnamefont {M.}~\bibnamefont {Weidem\"uller}},\ }\bibfield
  {title} {\bibinfo {title} {Observing the {D}ynamics of {D}ipole-{M}ediated
  {E}nergy {T}ransport by {I}nteraction-{E}nhanced {I}maging},\ }\href
  {https://doi.org/10.1126/science.1244843} {\bibfield  {journal} {\bibinfo
  {journal} {Science}\ }\textbf {\bibinfo {volume} {342}},\ \bibinfo {pages}
  {954} (\bibinfo {year} {2013})}\BibitemShut {NoStop}%
\bibitem [{\citenamefont {Schempp}\ \emph {et~al.}(2015)\citenamefont
  {Schempp}, \citenamefont {G\"unter}, \citenamefont {W\"uster}, \citenamefont
  {Weidem\"uller},\ and\ \citenamefont {Whitlock}}]{Schempp2015}%
  \BibitemOpen
  \bibfield  {author} {\bibinfo {author} {\bibfnamefont {H.}~\bibnamefont
  {Schempp}}, \bibinfo {author} {\bibfnamefont {G.}~\bibnamefont {G\"unter}},
  \bibinfo {author} {\bibfnamefont {S.}~\bibnamefont {W\"uster}}, \bibinfo
  {author} {\bibfnamefont {M.}~\bibnamefont {Weidem\"uller}},\ and\ \bibinfo
  {author} {\bibfnamefont {S.}~\bibnamefont {Whitlock}},\ }\bibfield  {title}
  {\bibinfo {title} {Correlated {E}xciton {T}ransport in
  {R}ydberg-{D}ressed-{A}tom {S}pin {C}hains},\ }\href
  {https://doi.org/10.1103/PhysRevLett.115.093002} {\bibfield  {journal}
  {\bibinfo  {journal} {Phys. Rev. Lett.}\ }\textbf {\bibinfo {volume} {115}},\
  \bibinfo {pages} {093002} (\bibinfo {year} {2015})}\BibitemShut {NoStop}%
\bibitem [{\citenamefont {Burioni}\ \emph {et~al.}(2000)\citenamefont
  {Burioni}, \citenamefont {Cassi}, \citenamefont {Meccoli}, \citenamefont
  {Rasetti}, \citenamefont {Regina}, \citenamefont {Sodano},\ and\
  \citenamefont {Vezzani}}]{Burioni2000}%
  \BibitemOpen
  \bibfield  {author} {\bibinfo {author} {\bibfnamefont {R.}~\bibnamefont
  {Burioni}}, \bibinfo {author} {\bibfnamefont {D.}~\bibnamefont {Cassi}},
  \bibinfo {author} {\bibfnamefont {I.}~\bibnamefont {Meccoli}}, \bibinfo
  {author} {\bibfnamefont {M.}~\bibnamefont {Rasetti}}, \bibinfo {author}
  {\bibfnamefont {S.}~\bibnamefont {Regina}}, \bibinfo {author} {\bibfnamefont
  {P.}~\bibnamefont {Sodano}},\ and\ \bibinfo {author} {\bibfnamefont
  {A.}~\bibnamefont {Vezzani}},\ }\bibfield  {title} {\bibinfo {title}
  {Bose-{E}instein condensation in inhomogeneous {J}osephson arrays},\ }\href
  {https://doi.org/10.1209/epl/i2000-00431-5} {\bibfield  {journal} {\bibinfo
  {journal} {Europhys. Lett.}\ }\textbf {\bibinfo {volume} {52}},\ \bibinfo
  {pages} {251} (\bibinfo {year} {2000})}\BibitemShut {NoStop}%
\bibitem [{\citenamefont {Burioni}\ \emph {et~al.}(2001)\citenamefont
  {Burioni}, \citenamefont {Cassi}, \citenamefont {Rasetti}, \citenamefont
  {Sodano},\ and\ \citenamefont {Vezzani}}]{Burioni2001}%
  \BibitemOpen
  \bibfield  {author} {\bibinfo {author} {\bibfnamefont {R.}~\bibnamefont
  {Burioni}}, \bibinfo {author} {\bibfnamefont {D.}~\bibnamefont {Cassi}},
  \bibinfo {author} {\bibfnamefont {M.}~\bibnamefont {Rasetti}}, \bibinfo
  {author} {\bibfnamefont {P.}~\bibnamefont {Sodano}},\ and\ \bibinfo {author}
  {\bibfnamefont {A.}~\bibnamefont {Vezzani}},\ }\bibfield  {title} {\bibinfo
  {title} {Bose-{E}instein condensation on inhomogeneous complex networks},\
  }\href
  {https://iopscience.iop.org/article/10.1088/0953-4075/34/23/314?pageTitle=IOPscience}
  {\bibfield  {journal} {\bibinfo  {journal} {J. Phys. B}\ }\textbf {\bibinfo
  {volume} {34}},\ \bibinfo {pages} {4697} (\bibinfo {year}
  {2001})}\BibitemShut {NoStop}%
\bibitem [{\citenamefont {Buonsante}\ \emph {et~al.}(2002)\citenamefont
  {Buonsante}, \citenamefont {Burioni}, \citenamefont {Cassi},\ and\
  \citenamefont {Vezzani}}]{Buonsante2002}%
  \BibitemOpen
  \bibfield  {author} {\bibinfo {author} {\bibfnamefont {P.}~\bibnamefont
  {Buonsante}}, \bibinfo {author} {\bibfnamefont {R.}~\bibnamefont {Burioni}},
  \bibinfo {author} {\bibfnamefont {D.}~\bibnamefont {Cassi}},\ and\ \bibinfo
  {author} {\bibfnamefont {A.}~\bibnamefont {Vezzani}},\ }\bibfield  {title}
  {\bibinfo {title} {Bose-{E}instein condensation on inhomogeneous networks:
  {M}esoscopic aspects versus thermodynamic limit},\ }\href
  {https://doi.org/10.1103/PhysRevB.66.094207} {\bibfield  {journal} {\bibinfo
  {journal} {Phys. Rev. B}\ }\textbf {\bibinfo {volume} {66}},\ \bibinfo
  {pages} {094207} (\bibinfo {year} {2002})}\BibitemShut {NoStop}%
\bibitem [{\citenamefont {Brunelli}\ \emph {et~al.}(2004)\citenamefont
  {Brunelli}, \citenamefont {Giusiano}, \citenamefont {Mancini}, \citenamefont
  {Sodano},\ and\ \citenamefont {Trombettoni}}]{Brunelli2004}%
  \BibitemOpen
  \bibfield  {author} {\bibinfo {author} {\bibfnamefont {I.}~\bibnamefont
  {Brunelli}}, \bibinfo {author} {\bibfnamefont {G.}~\bibnamefont {Giusiano}},
  \bibinfo {author} {\bibfnamefont {F.}~\bibnamefont {Mancini}}, \bibinfo
  {author} {\bibfnamefont {P.}~\bibnamefont {Sodano}},\ and\ \bibinfo {author}
  {\bibfnamefont {A.}~\bibnamefont {Trombettoni}},\ }\bibfield  {title}
  {\bibinfo {title} {Topology-induced spatial {B}ose-{E}instein condensation
  for bosons on star-shaped optical networks},\ }\href
  {https://iopscience.iop.org/article/10.1088/0953-4075/37/7/072} {\bibfield
  {journal} {\bibinfo  {journal} {J. Phys. B}\ }\textbf {\bibinfo {volume}
  {37}},\ \bibinfo {pages} {S275} (\bibinfo {year} {2004})}\BibitemShut
  {NoStop}%
\bibitem [{\citenamefont {Vidal}\ \emph {et~al.}(2011)\citenamefont {Vidal},
  \citenamefont {Lima},\ and\ \citenamefont {Lyra}}]{Vidal2011}%
  \BibitemOpen
  \bibfield  {author} {\bibinfo {author} {\bibfnamefont {E.~J. G.~G.}\
  \bibnamefont {Vidal}}, \bibinfo {author} {\bibfnamefont {R.~P.~A.}\
  \bibnamefont {Lima}},\ and\ \bibinfo {author} {\bibfnamefont {M.~L.}\
  \bibnamefont {Lyra}},\ }\bibfield  {title} {\bibinfo {title} {Bose-{E}instein
  condensation in the infinitely ramified star and wheel graphs},\ }\href
  {https://doi.org/10.1103/PhysRevE.83.061137} {\bibfield  {journal} {\bibinfo
  {journal} {Phys. Rev. E}\ }\textbf {\bibinfo {volume} {83}},\ \bibinfo
  {pages} {061137} (\bibinfo {year} {2011})}\BibitemShut {NoStop}%
\bibitem [{\citenamefont {de~Oliveira}\ \emph {et~al.}(2013)\citenamefont
  {de~Oliveira}, \citenamefont {dos Santos}, \citenamefont {de~Moura},
  \citenamefont {Lyra},\ and\ \citenamefont {Serva}}]{Oliveira2013}%
  \BibitemOpen
  \bibfield  {author} {\bibinfo {author} {\bibfnamefont {I.~N.}\ \bibnamefont
  {de~Oliveira}}, \bibinfo {author} {\bibfnamefont {T.~B.}\ \bibnamefont {dos
  Santos}}, \bibinfo {author} {\bibfnamefont {F.~A. B.~F.}\ \bibnamefont
  {de~Moura}}, \bibinfo {author} {\bibfnamefont {M.~L.}\ \bibnamefont {Lyra}},\
  and\ \bibinfo {author} {\bibfnamefont {M.}~\bibnamefont {Serva}},\ }\bibfield
   {title} {\bibinfo {title} {Critical behavior of the ideal-gas
  {B}ose-{E}instein condensation in the {A}pollonian network},\ }\href
  {https://doi.org/10.1103/PhysRevE.88.022139} {\bibfield  {journal} {\bibinfo
  {journal} {Phys. Rev. E}\ }\textbf {\bibinfo {volume} {88}},\ \bibinfo
  {pages} {022139} (\bibinfo {year} {2013})}\BibitemShut {NoStop}%
\bibitem [{\citenamefont {Lyra}\ \emph {et~al.}(2014)\citenamefont {Lyra},
  \citenamefont {de~Moura}, \citenamefont {de~Oliveira},\ and\ \citenamefont
  {Serva}}]{Lyra2014}%
  \BibitemOpen
  \bibfield  {author} {\bibinfo {author} {\bibfnamefont {M.~L.}\ \bibnamefont
  {Lyra}}, \bibinfo {author} {\bibfnamefont {F.~A. B.~F.}\ \bibnamefont
  {de~Moura}}, \bibinfo {author} {\bibfnamefont {I.~N.}\ \bibnamefont
  {de~Oliveira}},\ and\ \bibinfo {author} {\bibfnamefont {M.}~\bibnamefont
  {Serva}},\ }\bibfield  {title} {\bibinfo {title} {Bose-{E}instein
  condensation in diamond hierarchical lattices},\ }\href
  {https://doi.org/10.1103/PhysRevE.89.052133} {\bibfield  {journal} {\bibinfo
  {journal} {Phys. Rev. E}\ }\textbf {\bibinfo {volume} {89}},\ \bibinfo
  {pages} {052133} (\bibinfo {year} {2014})}\BibitemShut {NoStop}%
\bibitem [{\citenamefont {Buonsante}\ \emph {et~al.}(2004)\citenamefont
  {Buonsante}, \citenamefont {Burioni}, \citenamefont {Cassi}, \citenamefont
  {Penna},\ and\ \citenamefont {Vezzani}}]{Buonsante2004}%
  \BibitemOpen
  \bibfield  {author} {\bibinfo {author} {\bibfnamefont {P.}~\bibnamefont
  {Buonsante}}, \bibinfo {author} {\bibfnamefont {R.}~\bibnamefont {Burioni}},
  \bibinfo {author} {\bibfnamefont {D.}~\bibnamefont {Cassi}}, \bibinfo
  {author} {\bibfnamefont {V.}~\bibnamefont {Penna}},\ and\ \bibinfo {author}
  {\bibfnamefont {A.}~\bibnamefont {Vezzani}},\ }\bibfield  {title} {\bibinfo
  {title} {Topology-induced confined superfluidity in inhomogeneous arrays},\
  }\href {https://doi.org/10.1088/0953-4075/37/7/072} {\bibfield  {journal}
  {\bibinfo  {journal} {Phys. Rev. B}\ }\textbf {\bibinfo {volume} {70}},\
  \bibinfo {pages} {224510} (\bibinfo {year} {2004})}\BibitemShut {NoStop}%
\bibitem [{\citenamefont {Halu}\ \emph {et~al.}(2012)\citenamefont {Halu},
  \citenamefont {Ferretti}, \citenamefont {Vezzani},\ and\ \citenamefont
  {Bianconi}}]{Halu2012}%
  \BibitemOpen
  \bibfield  {author} {\bibinfo {author} {\bibfnamefont {A.}~\bibnamefont
  {Halu}}, \bibinfo {author} {\bibfnamefont {L.}~\bibnamefont {Ferretti}},
  \bibinfo {author} {\bibfnamefont {A.}~\bibnamefont {Vezzani}},\ and\ \bibinfo
  {author} {\bibfnamefont {G.}~\bibnamefont {Bianconi}},\ }\bibfield  {title}
  {\bibinfo {title} {Phase diagram of the {B}ose-{H}ubbard model on complex
  networks},\ }\href {https://doi.org/10.1209/0295-5075/99/18001} {\bibfield
  {journal} {\bibinfo  {journal} {Europhys. Lett.}\ }\textbf {\bibinfo {volume}
  {99}},\ \bibinfo {pages} {18001} (\bibinfo {year} {2012})}\BibitemShut
  {NoStop}%
\bibitem [{\citenamefont {DePue}\ \emph {et~al.}(1999)\citenamefont {DePue},
  \citenamefont {McCormick}, \citenamefont {Winoto}, \citenamefont {Oliver},\
  and\ \citenamefont {Weiss}}]{DePue1999}%
  \BibitemOpen
  \bibfield  {author} {\bibinfo {author} {\bibfnamefont {M.~T.}\ \bibnamefont
  {DePue}}, \bibinfo {author} {\bibfnamefont {C.}~\bibnamefont {McCormick}},
  \bibinfo {author} {\bibfnamefont {S.~L.}\ \bibnamefont {Winoto}}, \bibinfo
  {author} {\bibfnamefont {S.}~\bibnamefont {Oliver}},\ and\ \bibinfo {author}
  {\bibfnamefont {D.~S.}\ \bibnamefont {Weiss}},\ }\bibfield  {title} {\bibinfo
  {title} {Unity {O}ccupation of {S}ites in a 3{D} {O}ptical {L}attice},\
  }\href {https://doi.org/10.1103/PhysRevLett.82.2262} {\bibfield  {journal}
  {\bibinfo  {journal} {Phys. Rev. Lett.}\ }\textbf {\bibinfo {volume} {82}},\
  \bibinfo {pages} {2262} (\bibinfo {year} {1999})}\BibitemShut {NoStop}%
\bibitem [{\citenamefont {Bloch}\ \emph {et~al.}(2008)\citenamefont {Bloch},
  \citenamefont {Dalibard},\ and\ \citenamefont {Zwerger}}]{Bloch2008}%
  \BibitemOpen
  \bibfield  {author} {\bibinfo {author} {\bibfnamefont {I.}~\bibnamefont
  {Bloch}}, \bibinfo {author} {\bibfnamefont {J.}~\bibnamefont {Dalibard}},\
  and\ \bibinfo {author} {\bibfnamefont {W.}~\bibnamefont {Zwerger}},\
  }\bibfield  {title} {\bibinfo {title} {Many-body physics with ultracold
  gases},\ }\href {https://doi.org/10.1103/RevModPhys.80.885} {\bibfield
  {journal} {\bibinfo  {journal} {Rev. Mod. Phys.}\ }\textbf {\bibinfo {volume}
  {80}},\ \bibinfo {pages} {885} (\bibinfo {year} {2008})}\BibitemShut
  {NoStop}%
\bibitem [{\citenamefont {Chin}\ \emph {et~al.}(2010)\citenamefont {Chin},
  \citenamefont {Grimm}, \citenamefont {Julienne},\ and\ \citenamefont
  {Tiesinga}}]{Chin2010}%
  \BibitemOpen
  \bibfield  {author} {\bibinfo {author} {\bibfnamefont {C.}~\bibnamefont
  {Chin}}, \bibinfo {author} {\bibfnamefont {R.}~\bibnamefont {Grimm}},
  \bibinfo {author} {\bibfnamefont {P.}~\bibnamefont {Julienne}},\ and\
  \bibinfo {author} {\bibfnamefont {E.}~\bibnamefont {Tiesinga}},\ }\bibfield
  {title} {\bibinfo {title} {Feshbach resonances in ultracold gases},\ }\href
  {https://doi.org/10.1103/RevModPhys.82.1225} {\bibfield  {journal} {\bibinfo
  {journal} {Rev. Mod. Phys.}\ }\textbf {\bibinfo {volume} {82}},\ \bibinfo
  {pages} {1225} (\bibinfo {year} {2010})}\BibitemShut {NoStop}%
\bibitem [{\citenamefont {Weidem{\"u}ller}\ and\ \citenamefont
  {Zimmermann}(2011)}]{Weidemueller2011}%
  \BibitemOpen
  \bibfield  {author} {\bibinfo {author} {\bibfnamefont {M.}~\bibnamefont
  {Weidem{\"u}ller}}\ and\ \bibinfo {author} {\bibfnamefont {C.}~\bibnamefont
  {Zimmermann}},\ }\href {https://doi.org/10.1002/3527603417} {\emph {\bibinfo
  {title} {Interactions in ultracold gases: from atoms to molecules}}}\
  (\bibinfo  {publisher} {John Wiley \& Sons},\ \bibinfo {year}
  {2011})\BibitemShut {NoStop}%
\bibitem [{\citenamefont {Z\"urn}\ \emph {et~al.}(2012)\citenamefont {Z\"urn},
  \citenamefont {Serwane}, \citenamefont {Lompe}, \citenamefont {Wenz},
  \citenamefont {Ries}, \citenamefont {Bohn},\ and\ \citenamefont
  {Jochim}}]{Zuern2012}%
  \BibitemOpen
  \bibfield  {author} {\bibinfo {author} {\bibfnamefont {G.}~\bibnamefont
  {Z\"urn}}, \bibinfo {author} {\bibfnamefont {F.}~\bibnamefont {Serwane}},
  \bibinfo {author} {\bibfnamefont {T.}~\bibnamefont {Lompe}}, \bibinfo
  {author} {\bibfnamefont {A.~N.}\ \bibnamefont {Wenz}}, \bibinfo {author}
  {\bibfnamefont {M.~G.}\ \bibnamefont {Ries}}, \bibinfo {author}
  {\bibfnamefont {J.~E.}\ \bibnamefont {Bohn}},\ and\ \bibinfo {author}
  {\bibfnamefont {S.}~\bibnamefont {Jochim}},\ }\bibfield  {title} {\bibinfo
  {title} {Fermionization of two distinguishable fermions},\ }\href
  {https://doi.org/10.1103/PhysRevLett.108.075303} {\bibfield  {journal}
  {\bibinfo  {journal} {Phys. Rev. Lett.}\ }\textbf {\bibinfo {volume} {108}},\
  \bibinfo {pages} {075303} (\bibinfo {year} {2012})}\BibitemShut {NoStop}%
\bibitem [{\citenamefont {Tennie}\ \emph {et~al.}(2017)\citenamefont {Tennie},
  \citenamefont {Vedral},\ and\ \citenamefont {Schilling}}]{Tennie2017}%
  \BibitemOpen
  \bibfield  {author} {\bibinfo {author} {\bibfnamefont {F.}~\bibnamefont
  {Tennie}}, \bibinfo {author} {\bibfnamefont {V.}~\bibnamefont {Vedral}},\
  and\ \bibinfo {author} {\bibfnamefont {C.}~\bibnamefont {Schilling}},\
  }\bibfield  {title} {\bibinfo {title} {Universal upper bounds on the
  {B}ose-{E}instein condensate and the {H}ubbard star},\ }\href
  {https://doi.org/10.1103/PhysRevB.96.064502} {\bibfield  {journal} {\bibinfo
  {journal} {Phys. Rev. B}\ }\textbf {\bibinfo {volume} {96}},\ \bibinfo
  {pages} {064502} (\bibinfo {year} {2017})}\BibitemShut {NoStop}%
\bibitem [{\citenamefont {White}(1992)}]{White1992b}%
  \BibitemOpen
  \bibfield  {author} {\bibinfo {author} {\bibfnamefont {S.~R.}\ \bibnamefont
  {White}},\ }\bibfield  {title} {\bibinfo {title} {Density matrix formulation
  for quantum renormalization groups},\ }\href
  {https://doi.org/10.1103/PhysRevLett.69.2863} {\bibfield  {journal} {\bibinfo
   {journal} {Phys. Rev. Lett.}\ }\textbf {\bibinfo {volume} {69}},\ \bibinfo
  {pages} {2863} (\bibinfo {year} {1992})}\BibitemShut {NoStop}%
\bibitem [{\citenamefont {White}(1993)}]{White1993}%
  \BibitemOpen
  \bibfield  {author} {\bibinfo {author} {\bibfnamefont {S.~R.}\ \bibnamefont
  {White}},\ }\bibfield  {title} {\bibinfo {title} {Density-matrix algorithms
  for quantum renormalization groups},\ }\href
  {https://doi.org/10.1103/PhysRevB.48.10345} {\bibfield  {journal} {\bibinfo
  {journal} {Phys. Rev. B}\ }\textbf {\bibinfo {volume} {48}},\ \bibinfo
  {pages} {10345} (\bibinfo {year} {1993})}\BibitemShut {NoStop}%
\bibitem [{\citenamefont {Schollw\"ock}(2005)}]{Schollwock2005}%
  \BibitemOpen
  \bibfield  {author} {\bibinfo {author} {\bibfnamefont {U.}~\bibnamefont
  {Schollw\"ock}},\ }\bibfield  {title} {\bibinfo {title} {The density-matrix
  renormalization group},\ }\href {https://doi.org/10.1103/RevModPhys.77.259}
  {\bibfield  {journal} {\bibinfo  {journal} {Rev. Mod. Phys.}\ }\textbf
  {\bibinfo {volume} {77}},\ \bibinfo {pages} {259} (\bibinfo {year}
  {2005})}\BibitemShut {NoStop}%
\bibitem [{\citenamefont {Amico}\ \emph {et~al.}(2005)\citenamefont {Amico},
  \citenamefont {Osterloh},\ and\ \citenamefont {Cataliotti}}]{Amico2005}%
  \BibitemOpen
  \bibfield  {author} {\bibinfo {author} {\bibfnamefont {L.}~\bibnamefont
  {Amico}}, \bibinfo {author} {\bibfnamefont {A.}~\bibnamefont {Osterloh}},\
  and\ \bibinfo {author} {\bibfnamefont {F.}~\bibnamefont {Cataliotti}},\
  }\bibfield  {title} {\bibinfo {title} {Quantum {M}any {P}article {S}ystems in
  {R}ing-{S}haped {O}ptical {L}attices},\ }\href
  {https://doi.org/10.1103/PhysRevLett.95.063201} {\bibfield  {journal}
  {\bibinfo  {journal} {Phys. Rev. Lett.}\ }\textbf {\bibinfo {volume} {95}},\
  \bibinfo {pages} {063201} (\bibinfo {year} {2005})}\BibitemShut {NoStop}%
\bibitem [{\citenamefont {Franke-Arnold}\ \emph {et~al.}(2007)\citenamefont
  {Franke-Arnold}, \citenamefont {Leach}, \citenamefont {Padgett},
  \citenamefont {Lembessis}, \citenamefont {Ellinas}, \citenamefont {Wright},
  \citenamefont {Girkin}, \citenamefont {\"Ohberg},\ and\ \citenamefont
  {Arnold}}]{Franke2007}%
  \BibitemOpen
  \bibfield  {author} {\bibinfo {author} {\bibfnamefont {S.}~\bibnamefont
  {Franke-Arnold}}, \bibinfo {author} {\bibfnamefont {J.}~\bibnamefont
  {Leach}}, \bibinfo {author} {\bibfnamefont {M.~J.}\ \bibnamefont {Padgett}},
  \bibinfo {author} {\bibfnamefont {V.~E.}\ \bibnamefont {Lembessis}}, \bibinfo
  {author} {\bibfnamefont {D.}~\bibnamefont {Ellinas}}, \bibinfo {author}
  {\bibfnamefont {A.~J.}\ \bibnamefont {Wright}}, \bibinfo {author}
  {\bibfnamefont {J.~M.}\ \bibnamefont {Girkin}}, \bibinfo {author}
  {\bibfnamefont {P.}~\bibnamefont {\"Ohberg}},\ and\ \bibinfo {author}
  {\bibfnamefont {A.~S.}\ \bibnamefont {Arnold}},\ }\bibfield  {title}
  {\bibinfo {title} {Optical ferris wheel for ultracold atoms},\ }\href
  {https://doi.org/10.1364/OE.15.008619} {\bibfield  {journal} {\bibinfo
  {journal} {Opt. Express}\ }\textbf {\bibinfo {volume} {15}},\ \bibinfo
  {pages} {8619} (\bibinfo {year} {2007})}\BibitemShut {NoStop}%
\bibitem [{\citenamefont {Ramanathan}\ \emph {et~al.}(2011)\citenamefont
  {Ramanathan}, \citenamefont {Wright}, \citenamefont {Muniz}, \citenamefont
  {Zelan}, \citenamefont {Hill}, \citenamefont {Lobb}, \citenamefont
  {Helmerson}, \citenamefont {Phillips},\ and\ \citenamefont
  {Campbell}}]{Ramanathan2011}%
  \BibitemOpen
  \bibfield  {author} {\bibinfo {author} {\bibfnamefont {A.}~\bibnamefont
  {Ramanathan}}, \bibinfo {author} {\bibfnamefont {K.~C.}\ \bibnamefont
  {Wright}}, \bibinfo {author} {\bibfnamefont {S.~R.}\ \bibnamefont {Muniz}},
  \bibinfo {author} {\bibfnamefont {M.}~\bibnamefont {Zelan}}, \bibinfo
  {author} {\bibfnamefont {W.~T.}\ \bibnamefont {Hill}}, \bibinfo {author}
  {\bibfnamefont {C.~J.}\ \bibnamefont {Lobb}}, \bibinfo {author}
  {\bibfnamefont {K.}~\bibnamefont {Helmerson}}, \bibinfo {author}
  {\bibfnamefont {W.~D.}\ \bibnamefont {Phillips}},\ and\ \bibinfo {author}
  {\bibfnamefont {G.~K.}\ \bibnamefont {Campbell}},\ }\bibfield  {title}
  {\bibinfo {title} {Superflow in a toroidal {B}ose-{E}instein condensate: An
  atom circuit with a tunable weak link},\ }\href
  {https://doi.org/10.1103/PhysRevLett.106.130401} {\bibfield  {journal}
  {\bibinfo  {journal} {Phys. Rev. Lett.}\ }\textbf {\bibinfo {volume} {106}},\
  \bibinfo {pages} {130401} (\bibinfo {year} {2011})}\BibitemShut {NoStop}%
\bibitem [{\citenamefont {Amico}\ \emph {et~al.}(2014)\citenamefont {Amico},
  \citenamefont {Aghamalyan}, \citenamefont {Auksztol}, \citenamefont {Crepaz},
  \citenamefont {Dumke},\ and\ \citenamefont {Kwek}}]{Amico2014}%
  \BibitemOpen
  \bibfield  {author} {\bibinfo {author} {\bibfnamefont {L.}~\bibnamefont
  {Amico}}, \bibinfo {author} {\bibfnamefont {D.}~\bibnamefont {Aghamalyan}},
  \bibinfo {author} {\bibfnamefont {F.}~\bibnamefont {Auksztol}}, \bibinfo
  {author} {\bibfnamefont {H.}~\bibnamefont {Crepaz}}, \bibinfo {author}
  {\bibfnamefont {R.}~\bibnamefont {Dumke}},\ and\ \bibinfo {author}
  {\bibfnamefont {L.~C.}\ \bibnamefont {Kwek}},\ }\bibfield  {title} {\bibinfo
  {title} {Superfluid qubit systems with ring shaped optical lattices},\ }\href
  {https://doi.org/10.1038/srep04298} {\bibfield  {journal} {\bibinfo
  {journal} {Sci. Rep.}\ }\textbf {\bibinfo {volume} {4}},\ \bibinfo {pages}
  {4298} (\bibinfo {year} {2014})}\BibitemShut {NoStop}%
\bibitem [{\citenamefont {Bell}\ \emph {et~al.}(2016)\citenamefont {Bell},
  \citenamefont {Glidden}, \citenamefont {Humbert}, \citenamefont {Bromley},
  \citenamefont {Haine}, \citenamefont {Davis}, \citenamefont {Neely},
  \citenamefont {Baker},\ and\ \citenamefont {Rubinsztein-Dunlop}}]{Bell2016}%
  \BibitemOpen
  \bibfield  {author} {\bibinfo {author} {\bibfnamefont {T.~A.}\ \bibnamefont
  {Bell}}, \bibinfo {author} {\bibfnamefont {J.~A.~P.}\ \bibnamefont
  {Glidden}}, \bibinfo {author} {\bibfnamefont {L.}~\bibnamefont {Humbert}},
  \bibinfo {author} {\bibfnamefont {M.~W.~J.}\ \bibnamefont {Bromley}},
  \bibinfo {author} {\bibfnamefont {S.~A.}\ \bibnamefont {Haine}}, \bibinfo
  {author} {\bibfnamefont {M.~J.}\ \bibnamefont {Davis}}, \bibinfo {author}
  {\bibfnamefont {T.~W.}\ \bibnamefont {Neely}}, \bibinfo {author}
  {\bibfnamefont {M.~A.}\ \bibnamefont {Baker}},\ and\ \bibinfo {author}
  {\bibfnamefont {H.}~\bibnamefont {Rubinsztein-Dunlop}},\ }\bibfield  {title}
  {\bibinfo {title} {Bose-{E}instein condensation in large time-averaged
  optical ring potentials},\ }\href
  {https://doi.org/10.1088/1367-2630/18/3/035003} {\bibfield  {journal}
  {\bibinfo  {journal} {New. J. Phys.}\ }\textbf {\bibinfo {volume} {18}},\
  \bibinfo {pages} {035003} (\bibinfo {year} {2016})}\BibitemShut {NoStop}%
\bibitem [{\citenamefont {Sakmann}\ \emph {et~al.}(2008)\citenamefont
  {Sakmann}, \citenamefont {Streltsov}, \citenamefont {Alon},\ and\
  \citenamefont {Cederbaum}}]{Sakmann2008}%
  \BibitemOpen
  \bibfield  {author} {\bibinfo {author} {\bibfnamefont {K.}~\bibnamefont
  {Sakmann}}, \bibinfo {author} {\bibfnamefont {A.~I.}\ \bibnamefont
  {Streltsov}}, \bibinfo {author} {\bibfnamefont {O.~E.}\ \bibnamefont
  {Alon}},\ and\ \bibinfo {author} {\bibfnamefont {L.~S.}\ \bibnamefont
  {Cederbaum}},\ }\bibfield  {title} {\bibinfo {title} {Reduced density
  matrices and coherence of trapped interacting bosons},\ }\href
  {https://doi.org/???} {\bibfield  {journal} {\bibinfo  {journal} {Phys. Rev.
  A}\ }\textbf {\bibinfo {volume} {78}},\ \bibinfo {pages} {023615} (\bibinfo
  {year} {2008})}\BibitemShut {NoStop}%
\bibitem [{\citenamefont {Legeza}\ and\ \citenamefont
  {S\'olyom}(2004)}]{Legeza-2004b}%
  \BibitemOpen
  \bibfield  {author} {\bibinfo {author} {\bibfnamefont {O.}~\bibnamefont
  {Legeza}}\ and\ \bibinfo {author} {\bibfnamefont {J.}~\bibnamefont
  {S\'olyom}},\ }\bibfield  {title} {\bibinfo {title} {Quantum data
  compression, quantum information generation, and the density-matrix
  renormalization-group method},\ }\href
  {https://doi.org/10.1103/PhysRevB.70.205118} {\bibfield  {journal} {\bibinfo
  {journal} {Phys. Rev. B}\ }\textbf {\bibinfo {volume} {70}},\ \bibinfo
  {pages} {205118} (\bibinfo {year} {2004})}\BibitemShut {NoStop}%
\bibitem [{\citenamefont {Rissler}\ \emph {et~al.}(2006)\citenamefont
  {Rissler}, \citenamefont {Noack},\ and\ \citenamefont {White}}]{Rissler2006}%
  \BibitemOpen
  \bibfield  {author} {\bibinfo {author} {\bibfnamefont {J.}~\bibnamefont
  {Rissler}}, \bibinfo {author} {\bibfnamefont {R.~M.}\ \bibnamefont {Noack}},\
  and\ \bibinfo {author} {\bibfnamefont {S.~R.}\ \bibnamefont {White}},\
  }\bibfield  {title} {\bibinfo {title} {Measuring orbital interaction using
  quantum information theory},\ }\href
  {https://doi.org/http://dx.doi.org/10.1016/j.chemphys.2005.10.018} {\bibfield
   {journal} {\bibinfo  {journal} {Chemical Physics}\ }\textbf {\bibinfo
  {volume} {323}},\ \bibinfo {pages} {519 } (\bibinfo {year}
  {2006})}\BibitemShut {NoStop}%
\bibitem [{\citenamefont {Szalay}\ \emph {et~al.}(2015)\citenamefont {Szalay},
  \citenamefont {Pfeffer}, \citenamefont {Murg}, \citenamefont {Barcza},
  \citenamefont {Verstraete}, \citenamefont {Schneider},\ and\ \citenamefont
  {Legeza}}]{Szalay2015}%
  \BibitemOpen
  \bibfield  {author} {\bibinfo {author} {\bibfnamefont {S.}~\bibnamefont
  {Szalay}}, \bibinfo {author} {\bibfnamefont {M.}~\bibnamefont {Pfeffer}},
  \bibinfo {author} {\bibfnamefont {V.}~\bibnamefont {Murg}}, \bibinfo {author}
  {\bibfnamefont {G.}~\bibnamefont {Barcza}}, \bibinfo {author} {\bibfnamefont
  {F.}~\bibnamefont {Verstraete}}, \bibinfo {author} {\bibfnamefont
  {R.}~\bibnamefont {Schneider}},\ and\ \bibinfo {author} {\bibfnamefont
  {O.}~\bibnamefont {Legeza}},\ }\bibfield  {title} {\bibinfo {title} {Tensor
  product methods and entanglement optimization for ab initio quantum
  chemistry},\ }\href
  {https://onlinelibrary.wiley.com/doi/abs/10.1002/qua.24898} {\bibfield
  {journal} {\bibinfo  {journal} {Int. J. Quant. Chem.}\ }\textbf {\bibinfo
  {volume} {115}},\ \bibinfo {pages} {1342} (\bibinfo {year}
  {2015})}\BibitemShut {NoStop}%
\bibitem [{\citenamefont {Bethe}(1931)}]{Bethe1931}%
  \BibitemOpen
  \bibfield  {author} {\bibinfo {author} {\bibfnamefont {H.}~\bibnamefont
  {Bethe}},\ }\bibfield  {title} {\bibinfo {title} {Zur {T}heorie der
  {M}etalle},\ }\href {https://doi.org/???} {\bibfield  {journal} {\bibinfo
  {journal} {Z. Phys.}\ }\textbf {\bibinfo {volume} {71}},\ \bibinfo {pages}
  {1703} (\bibinfo {year} {1931})}\BibitemShut {NoStop}%
\bibitem [{\citenamefont {Nie}\ \emph {et~al.}(2013)\citenamefont {Nie},
  \citenamefont {Katsura},\ and\ \citenamefont {Oshikawa}}]{Nie2013}%
  \BibitemOpen
  \bibfield  {author} {\bibinfo {author} {\bibfnamefont {W.}~\bibnamefont
  {Nie}}, \bibinfo {author} {\bibfnamefont {H.}~\bibnamefont {Katsura}},\ and\
  \bibinfo {author} {\bibfnamefont {M.}~\bibnamefont {Oshikawa}},\ }\bibfield
  {title} {\bibinfo {title} {Ground {S}tate {E}nergies of {S}pinless {F}ree
  {F}ermions and {H}ard {C}ore {B}osons},\ }\href {https://doi.org/???}
  {\bibfield  {journal} {\bibinfo  {journal} {Phys. Rev. Lett.}\ }\textbf
  {\bibinfo {volume} {111}},\ \bibinfo {pages} {100402} (\bibinfo {year}
  {2013})}\BibitemShut {NoStop}%
\bibitem [{\citenamefont {Nie}\ \emph {et~al.}(2018)\citenamefont {Nie},
  \citenamefont {Katsura},\ and\ \citenamefont {Oshikawa}}]{Nie2018}%
  \BibitemOpen
  \bibfield  {author} {\bibinfo {author} {\bibfnamefont {W.}~\bibnamefont
  {Nie}}, \bibinfo {author} {\bibfnamefont {H.}~\bibnamefont {Katsura}},\ and\
  \bibinfo {author} {\bibfnamefont {M.}~\bibnamefont {Oshikawa}},\ }\bibfield
  {title} {\bibinfo {title} {Particle statistics, frustration, and ground state
  energy},\ }\href {https://doi.org/???} {\bibfield  {journal} {\bibinfo
  {journal} {Phys. Rev. B}\ }\textbf {\bibinfo {volume} {97}},\ \bibinfo
  {pages} {125153} (\bibinfo {year} {2018})}\BibitemShut {NoStop}%
\bibitem [{\citenamefont {Pethick}\ and\ \citenamefont
  {Pitaevskii}(2000)}]{Pethick2000}%
  \BibitemOpen
  \bibfield  {author} {\bibinfo {author} {\bibfnamefont {C.~J.}\ \bibnamefont
  {Pethick}}\ and\ \bibinfo {author} {\bibfnamefont {L.~P.}\ \bibnamefont
  {Pitaevskii}},\ }\bibfield  {title} {\bibinfo {title} {Criterion for
  {B}ose-{E}instein condensation for particles in traps},\ }\href
  {https://doi.org/???} {\bibfield  {journal} {\bibinfo  {journal} {Phys. Rev.
  A}\ }\textbf {\bibinfo {volume} {62}},\ \bibinfo {pages} {033609} (\bibinfo
  {year} {2000})}\BibitemShut {NoStop}%
\end{thebibliography}%

\onecolumngrid
\newpage

\appendix

\setcounter{equation}{0}
\setcounter{figure}{0}
\setcounter{table}{0}
\makeatletter
\renewcommand{\theequation}{S\arabic{equation}}
\renewcommand{\thefigure}{S\arabic{figure}}

\section*{Supplementary Methods}

\subsection*{Derivation of the effective Hamiltonian}\label{app:effective-hamiltonian}

The Hamiltonian $\hat{H}$ is given by Eq.~(1). Using  $\hat{H} = \hat{H}_0 + \hat{H}_1$ (with $\hat{H}_0$ the ring-Hamiltonian) and

\begin{equation}\label{eq:Psi-Suppl}
\ket{\Psi_N} = \alpha \ket{\phi_N}\otimes \ket{0}_c + \beta \ket{\varphi_{N-1}}\otimes \ket{1}_c \ ,
\end{equation}
the eigenvalue equation $\hat{H} \ket{\Psi_N} = E\ket{\Psi_N} $ becomes
\begin{eqnarray}\label{eq:eigenvalue-1A}
\alpha \hat{H}_0 \ket{\phi_N} - \beta s \sum^d_{i=1} h^{\dagger}_i  \ket{\varphi_{N-1}} &=& \alpha E \ket{\phi_N}\nonumber \\
 -\alpha s \sum^d_{i=1} h_i  \ket{\phi_{N}} + \beta H_0\ket{\varphi_{N-1}} &=& \beta E  \ket{\varphi_{N-1}} \ .
\end{eqnarray}
The unperturbed eigenstates (i.e., s=0) for $N$ HCBs can be represented as
\begin{equation}\label{eq:psi-0-0}
\ket{\psi^0_{\boldsymbol{\mu}}(N)} = \sum_{1 \leq n_1< \cdots < n_N \leq d} \psi^0_{\boldsymbol{\mu}}(n_1,\cdots,n_N)h_{n_1}^{\dagger}  \cdots h_{n_N}^{\dagger} \ket{0}   \ .
\end{equation}
The normalized, totally symmetric `wave functions' $\psi^0_{\boldsymbol{\mu}}(n_1,\cdots,n_N)$ are given for $1 \leq n_1< \cdots < n_N \leq d$ by  the determinant constructed from  the one-particle states $\exp(iq_{\mu_k}n_l)$ \cite{Bethe1931,Lieb1963a}
\begin{equation}\label{eq:psi-0}
 \psi^0_{\boldsymbol{\mu}}(n_1,\cdots,n_N) = \mathcal{N}  \sum_{P \in S_N} sgn(P)
 \exp{\big(i \sum^N_{k=1}q_{\mu_{P(k)}}n_k\big)}  \ ,
\end{equation}
with $\mathcal{N} = d^{-N/2}$. $S_N$ denotes the  permutation group of the integers $(1,2,\cdots,N)$ and $sgn(P)$ its signature.
The form \eqref{eq:psi-0} highlights the well-known equivalence of spinless fermions and hard-core bosons in 1d and one has $\mu_1 < \mu_2< \ldots <\mu_d$.
The unperturbed eigenstates are labelled by  $\boldsymbol{\mu}=(\mu_1,\cdots,\mu_N)$
and $\mu$ determines the wave number $q_{\mu}= (\pi/d)(2\mu+1)$ for $N$ even and  $q_{\mu}= (\pi/d)2 \mu$ for
$N$ odd \cite{Lieb1963a,Nie2013,Nie2018}
.  Since $q_{\mu}$ is restricted to the first Billouin zone $\mu$ takes the values
$-d/2+1,-d/2+2,\cdots,-1,0,1,\cdots,d/2-1,d/2 $ for $d$ even and $-(d-1)/2,-(d-1)/2+1,\cdots,-1,0,1,\cdots,(d-1)/2-1,(d-1)/2 $ for $d$ odd.  The corresponding unperturbed eigenvalues are given by
\begin{equation}\label{eq:E-0}
 E^{0}_{\boldsymbol{\mu}}(N,d) = -2 \sum^N_{k=1} \cos(q_{\mu_k})\ .
\end{equation}
The unperturbed ground state energy, $E^{0}_{low}(N,d)$,  is easily calculated. Using  $\cos(x)= [\exp{(ix)} + \exp{(-ix)}]/2$ it is straightforward to calculate the  sum in Eq.~(\ref{eq:E-0}). As a result one gets
\begin{equation}\label{eq:E-0-a}
 E^{0}_{low}(N,d) = -2 \sin(\frac{\pi}{d}N)/\sin(\frac{\pi}{d}) \simeq   -2 \frac{d}{\pi}\sin(\frac{\pi}{d}N)    \ ,
\end{equation}
and the Fermi energy  $E_F(N,d)= E^0_0(N,d) - E^0_0(N-1,d)$ becomes
\begin{equation}\label{eq:E-Fermi}
 E_F(N,d) = -2 \big[\tan(\frac{\pi}{2d})\sin(\frac{\pi}{d}N)+\cos(\frac{\pi}{d}N) \big] \simeq -2\cos(\pi n)  \
\end{equation}
with $n=N/d$ the particle density.  For $d \to \infty, N \to \infty$ with $n=N/d$ fixed the unperturbed $N$-particle spectrum is a single band with lower  band edge  $E^0_{low}(N,n) \simeq -2N \sin(\pi n) /(\pi n)$
and band width $W(N,n) = 4N \sin(\pi n) /(\pi n) \propto 4N$.
First we note that the result for $E^{0}_{low}(N,n)$ holds for N even and odd
and second that the ground state lies in the subspace with total momentum  $Q=\sum^N_{k=1} q_{\mu_k}$ equal to zero. $Q$ is a good  quantum number  due to the invariance of  $\hat{H}$  under lattice translations on the ring.

Since the unperturbed eigenstates are complete we have(cf. also Eq.~(3) of the main text)
\begin{eqnarray}\label{eq:phi-varphi}
\ket{\phi_N} &=& \sum_{\boldsymbol{\nu}} A_{\boldsymbol{\nu}}\ket{\psi^0_{\boldsymbol{\nu}}(N)} \nonumber \\
\ket{\varphi_{N-1}} &=& \sum_{\boldsymbol{\mu}} a_{\boldsymbol{\mu}}\ket{\psi^0_{\boldsymbol{\mu}}(N-1)}\ .
\end{eqnarray}
with $\sum_{\boldsymbol{\nu}} |A_{\boldsymbol{\nu}}|^2 =1$ and $\sum_{\boldsymbol{\mu}} |a_{\boldsymbol{\mu}}|^2 =1$.  The summations in Eq.~(\ref{eq:phi-varphi}) are restricted such that  $Q= \sum^{N}_{i=1} q_{\nu_i}=\sum^{N-1}_{j=1} q_{\mu_j}$ is fixed (mod $2\pi$). In the following we choose $Q=0$, i.e., we consider the HCBs in the  frame where the center of mass of the HCBs is at rest(see also \cite{Pethick2000}).
Substituting the ansatz (\ref{eq:phi-varphi}) into Eq.~(\ref{eq:eigenvalue-1A}) leads to a decoupling of the $(N-1)$-particle and the $N$-particle sector:
\begin{eqnarray}\label{eq:eigenvalue-2A}
  \big[E-E^0_{\boldsymbol{\nu}}(N,d) \big] A_{\boldsymbol{\nu}} &=& s^2\sum_{\boldsymbol{\nu}'}M_{\boldsymbol{\nu}\boldsymbol{\nu}'}(E)A_{\boldsymbol{\nu}'} \nonumber \\
\big[E-E^0_{\boldsymbol{\mu}}(N-1,d) \big]a_{\boldsymbol{\mu}} &=& s^2\sum_{\boldsymbol{\mu}'}m_{\boldsymbol{\mu}\boldsymbol{\mu}'}(E)a_{\boldsymbol{\mu}'} \  .
\end{eqnarray}
Here we used that $E^0_{\boldsymbol{\mu}}(N-1,d)$ and $E^0_{\boldsymbol{\nu}}(N,d)$ are the corresponding unperturbed eigenvalues of
$\ket{\psi^0_{\boldsymbol{\mu}}(N-1)}$ and  $\ket{\psi^0_{\boldsymbol{\nu}}(N)}$
 , respectively. The matrix elements  $M_{\boldsymbol{\nu}\boldsymbol{\nu}'}(E)$ and $m_{\boldsymbol{\mu}\boldsymbol{\mu}'}(E)$ depend only on the unperturbed eigenstates and eigenvalues and are given by
\begin{eqnarray}\label{eq:eigenvalue-3A}
  M_{\boldsymbol{\nu}\boldsymbol{\nu}'}(E) &=& \sum_{\boldsymbol{\mu}'}(b^{\dagger})_{\boldsymbol{\nu}\boldsymbol{\mu}'} [E-E^0_{\boldsymbol{\mu}'}(N-1,d)]^{-1}b_{\boldsymbol{\mu}'\boldsymbol{\nu}'} \nonumber \\
 m_{\boldsymbol{\mu}\boldsymbol{\mu}'}(E) &=& \sum_{\boldsymbol{\nu}'} b_{\boldsymbol{\mu}\boldsymbol{\nu}'} [E-E^0_{\boldsymbol{\nu}'}(N,d)]^{-1}b^{\dagger}_{\boldsymbol{\nu}'\boldsymbol{\mu}'} \  .
\end{eqnarray}
The crucial quantity is the matrix $\bf{b}$ with elements
\begin{equation}\label{eq:eigenvalue-4A}
b_{\boldsymbol{\mu}\boldsymbol{\nu}} = \bra{\psi^0_{\boldsymbol{\mu}}(N-1)} \ \sum^{d}_{i=1}h_i \ \ket{\psi^0_{\boldsymbol{\nu}}(N)}  \  \ .
\end{equation}
 Having solved Eq.~(\ref{eq:eigenvalue-2A}) one obtains  from Eq.~(\ref{eq:eigenvalue-1A}) with Eq.~(\ref{eq:phi-varphi}) the coefficients $\alpha$ and $\beta$.

Operating with $\sum_{\boldsymbol{\nu}}A^{*}_{\boldsymbol{\nu}} [E-E^0_{\boldsymbol{\nu}}(N,d)]^{-1}$ and $\sum_{\boldsymbol{\mu}}a^{*}_{\boldsymbol{\mu}} [E-E^0_{\boldsymbol{\mu}}(N-1,d)]^{-1} $, respectively, on the 1st and 2nd line  of Eq.~(\ref{eq:eigenvalue-2A}) and taking the normalization of $\{A_{\boldsymbol{\nu}}\}$ and
$\{a_{\boldsymbol{\mu}}\}$ into account the eigenvalue equations  take the form
\begin{eqnarray}\label{eq:eigenvalue-5A}
1&=&s^2f^{(N-1)}_d(E;\{a_{\boldsymbol{\mu}}\})
 \nonumber \\
 1&=&s^2F^{(N)}_d(E;\{A_{\boldsymbol{\nu}}\}) \ ,
\end{eqnarray}
with
\begin{equation}\label{eq:eigenvalue-6Ai}
 f^{(N-1)}_d(E;\{a_{\boldsymbol{\mu}}\}) =\sum_{\boldsymbol{\mu}\boldsymbol{\mu}'\boldsymbol{\nu}'}a_{\boldsymbol{\mu}}^{*} [E-E^0_{\boldsymbol{\mu}}(N-1,d)]^{-1}  b_{\boldsymbol{\mu}\boldsymbol{\nu}'} [E-E^0_{\boldsymbol{\nu}'}(N,d)]^{-1}(b^{\dagger})_{\boldsymbol{\nu}'\boldsymbol{\mu}'}a_{\boldsymbol{\mu}'}  \ .
\end{equation}
and
\begin{equation}\label{eq:eigenvalue-6Aii}
 F^{(N)}_d(E;\{A_{\boldsymbol{\nu}}\}) =\sum_{\boldsymbol{\nu}\boldsymbol{\nu}'\boldsymbol{\mu}'}A_{\boldsymbol{\nu}}^{*} [E-E^0_{\boldsymbol{\nu}}(N,d)]^{-1} (b^{\dagger})_{\boldsymbol{\nu}\boldsymbol{\mu}'} [E-E^0_{\boldsymbol{\mu}'}(N-1,d)]^{-1}b_{\boldsymbol{\mu}'\boldsymbol{\nu}'}A_{\boldsymbol{\nu}'}
\end{equation}
The unperturbed eigenfunctions  can always be chosen to be real, since the unperturbed Hamiltonian is real. Therefore $b_{\boldsymbol{\mu}\boldsymbol{\nu}}$ and $(b^{\dagger})_{\boldsymbol{\nu}\boldsymbol{\mu}}$ are real. Furthermore, $\{ a_{\boldsymbol{\mu}} \}$
and $\{ A_{\boldsymbol{\nu}}  \}$ can also be chosen
to be real since the Hamiltonian $\hat{H}$ is real as well. Therefore the functions $ f^{(N-1)}_d(E;\{a_{\boldsymbol{\mu}}\})$
and $ F^{(N)}_d(E;\{A_{\boldsymbol{\nu}}\})$ are real.
\\

Eq.~(\ref{eq:eigenvalue-5A}) together with Eqs.~(\ref{eq:eigenvalue-6Ai}) and (\ref{eq:eigenvalue-6Aii}) already allows to obtain some qualitative  information on the low-energy part of the perturbed $N$-particle spectrum.
A crucial observation is  that
 $f^{(N-1)}_d(E;\{a_{\boldsymbol{\mu}}\})$ and  $F^{(N)}_d(E;\{A_{\boldsymbol{\nu}}\})$ have \textit{poles} at the unperturbed $(N-1)$- and $N$-particle eigenvalues.
As discussed above  the unperturbed spectrum of $(N-1)$  and $N$ HCBs form a band with lower band edge $E^{0}_{low}(N-1,d)$
 and $E^{0}_{low}(N,d)$, respectively. It is  $E^{0}_{low}(N,d) = E^{0}_{low}(N-1,d) + E_F(N,d)$ with the Fermi energy from Eq.~(\ref{eq:E-Fermi}). For $n < 1/2$ it follows $E_F(N,d) < 0$. Note, this is not a restriction due to the particle-hole duality. Therefore,
 $E^{0}_{low}(N,d) < E^{0}_{low}(N-1,d)$. Now, we  will show that  the low-lying perturbed eigenvalues
form a discrete spectrum below $E^{0}_{low}(N,d)$, and the band between $E^{0}_{low}(N,d)$ and $E^{0}_{low}(N-1,d)$ persists. The discrete spectrum exhibits an excitation gap even for $d= \infty$.

Let us choose $E$ between $E^0_{low}(N,d)$ and $E^0_{low}(N-1,d)$ and let us denote  the increasingly ordered unperturbed eigenvalues $\{ E^0_{\boldsymbol{\nu}}(N,d) \}$ in this interval  by $E^0_{\boldsymbol{\nu}_k}(N,d)$, $k \geq 1 $.  Due to $E < E^{0}_{low}(N-1,d)$
the denominators $ [E-E^0_{\boldsymbol{\mu}}(N-1,d)]^{-1}$ in Eq.~(\ref{eq:eigenvalue-6Ai}) are negative for all
$\boldsymbol{\mu}$, i.e., they do not change sign. Then, under variation of $E$ between $E^0_{\boldsymbol{\nu}_k}(N,d)$ and $E^0_{\boldsymbol{\nu}_{k+1}}(N,d)$, the function
$f^{(N-1)}_d(E;\{a_{\boldsymbol{\mu}}\})$ varies continuously from $\pm \infty$ at $E=E^0_{\boldsymbol{\nu}_k}(N,d)$ to  $\mp \infty$ at $E=E^0_{\boldsymbol{\nu}_{k+1}}(N,d)$, independent of   $\{a_{\boldsymbol{\mu}}\}$.  Accordingly, for arbitrary $s \neq 0$ the first equation of (\ref{eq:eigenvalue-5A}) has always  a solution $E_{\boldsymbol{\nu}_k}(N, d; s, \{a_{\boldsymbol{\nu}}\})$ which is between $E^{0}_{\boldsymbol{\nu}_k}(N,d)$ and
$E^{0}_{\boldsymbol{\nu}_{k+1}}(N,d)$.
Substituting  $E_{\boldsymbol{\nu}_k}(N, d; s, \{a_{\boldsymbol{\nu}}\})$ into the second line of Eq.~(\ref{eq:eigenvalue-2A})
yields $\{a^{({\boldsymbol{\nu}_k})}_{\boldsymbol{\nu}}\}$ which in turn leads to  the perturbed eigenvalues $E_{\boldsymbol{\nu}_k}(N,d;s)= E_{\boldsymbol{\nu}_k}(N, d; s, \{a^{({\boldsymbol{\nu}_k})}_{\boldsymbol{\nu}}\}) $, $k \geq 1$.  For $d \to \infty$ these perturbed eigenvalues $\{E_{\boldsymbol{\nu}_k}(N,d; s) \}$ form a band with lower band edge $E^{0}_{low}(N,d)$ and upper edge $E^{0}_{low}(N-1,d)$.

For  $E \geq E^{0}_{low}(N-1,d)$   there exist pairs $\boldsymbol{\nu}', \boldsymbol{\mu}$  such that there is no unperturbed eigenvalue between $E^0_{\boldsymbol{\nu}'}(N,d)$ and $E^0_{\boldsymbol{\mu}}(N-1,d)$. In that case
$ [E-E^0_{\boldsymbol{\nu}'}(N,d)]^{-1} [E-E^0_{\boldsymbol{\mu}}(N-1,d)]^{-1}$ in Eq.~(\ref{eq:eigenvalue-6Ai})
changes from $\pm \infty$ at $E^0_{\boldsymbol{\nu}'}(N,d)$ to $\pm \infty$  at $E^0_{\boldsymbol{\mu}}(N-1,d)$ under varying $E$  between $E^0_{\boldsymbol{\nu}'}(N,d)$ and $E^0_{\boldsymbol{\mu}}(N-1,d)$. Therefore, $ f^{(N-1)}_d(E;\{a_{\boldsymbol{\nu}}\})$ does not necessarily change sign and the first equation of (\ref{eq:eigenvalue-5A}) may only have a solution for  $s^2$ small enough. In case that
the solution between $E^0_{\boldsymbol{\nu}'}(N,d)$ and $E^0_{\boldsymbol{\mu}}(N-1,d)$ disappears if $s^2$ becomes large enough, a perturbed eigenvalue must appear below(or above)  the  lower(upper) band edge $-|E^{0}_{low}(N,d)|$($|E^{0}_{low}(N-1,d)|$), since the total number of eigenvalues does not depend on $s$.

Finally, let us discuss $E < E^{0}_{low}(N,d)$. In that case the  product of both denominators in  Eq.~(\ref{eq:eigenvalue-6Ai}) is always positive. For $E \to E^{0}_{low}(N,d)$ from below  $f^{(N-1)}_d(E;\{a_{\boldsymbol{\mu}}\})$ will diverge to $ + \infty$.
Because $f^{(N-1)}_d(E;\{a_{\boldsymbol{\mu}}\}) \to 0$ for $E \to  - \infty$  there must exist at least one solution
$E(N,d;s,\{a_{\boldsymbol{\mu}}\})$ of the first equation of Eq.~(\ref{eq:eigenvalue-5A}) for \textit{all} $s \neq 0$. $E(N,d;s,\{a_{\boldsymbol{\mu}}\})$ will have a gap to the lower band edge $E^{0}_{low}(N,d)$. Depending on $s^2$ and $\{a_{\boldsymbol{\mu}}\}$ there may exist more than one solution. Substitution them into the first line of Eq.~(\ref{eq:eigenvalue-2A}) yields a discrete spectrum. The same qualitative discussion can be done for $F^{(N)}_d(E;\{A_{\boldsymbol{\nu}}\}) $
(Eq.~(\ref{eq:eigenvalue-6Aii})) in combination with the second equation in (\ref{eq:eigenvalue-5A}).
We have checked the correctness of these qualitative results on the perturbed spectrum for $N=2$. There is little doubt that they become incorrect for $N > 2$.
\\
\\

Now we describe how the discrete part of the perturbed spectrum  and the corresponding eigenstates below $E^0_{low}(N,d)$ can be obtained exactly in two limiting cases. In these two cases  $ [E-E^0_{\boldsymbol{\mu'}}(N-1,d)]^{-1}$ and  $ [E-E^0_{\boldsymbol{\nu}'}(N,d)]^{-1}$ can be replaced by $ [E-E^0_{low}(N,d) + E_F(N,d)]^{-1}$ and  $ [E-E^0_{low}(N,d)]^{-1}$, respectively.  Then, the matrices   $ (M_{\boldsymbol{\nu}\boldsymbol{\nu}'}(E))$ and $ (m_{\boldsymbol{\mu}\boldsymbol{\mu}'}(E))$ strongly simplify since the sums in Eq.~(\ref{eq:eigenvalue-3A})  can be performed using
Eq.~(\ref{eq:eigenvalue-4A}) and the completeness relations $\sum_{\boldsymbol{\mu}} \ket{\psi^0_{\boldsymbol{\mu}}(N-1)} \bra{\psi^0_{\boldsymbol{\mu}}(N-1)} =$ $\mathbbm{1}|_{N-1}$, \  $\sum_{\boldsymbol{\nu}} \ket{\psi^0_{\boldsymbol{\nu}}(N)} \bra{\psi^0_{\boldsymbol{\nu}}(N)} =$ $\mathbbm{1}|_N$.   $\mathbbm{1}|_{N-1}$ and  $\mathbbm{1}|_{N}$ denote the identity operator, respectively,  in the $(N-1)$ and $N$ particle subspace. Then it follows
\begin{eqnarray}\label{eq:eigenvalue-6Aiiapprox}
M_{\boldsymbol{\nu}\boldsymbol{\nu}'}(E) &\simeq&
   [E-E^0_{low}(N,d) + E_F(N,d)]^{-1} \bra{\psi^0_{\boldsymbol{\nu}}(N)} \sum_{ij}h^{\dagger}_ih_j \ket{\psi^0_{\boldsymbol{\nu'}}(N)} \nonumber \\
 m_{\boldsymbol{\mu}\boldsymbol{\mu}'}(E) &\simeq &  [E-E^0_{low}(N,d)]^{-1} \bra{\psi^0_{\boldsymbol{\mu}}(N-1)} \sum_{ij}h_ih^{\dagger}_j \ket{\psi^0_{\boldsymbol{\mu'}}(N-1)} \nonumber\  .
\end{eqnarray}
and Eq.~(\ref{eq:eigenvalue-2A}) simplifies to
\begin{eqnarray}\label{eq:eigenvalue-6Aiiiapprox}
E^{eff} A_{\boldsymbol{\nu}} & \simeq& s^2\sum_{\boldsymbol{\nu}'} \bra{\psi^0_{\boldsymbol{\nu}}(N)} \sum_{ij}h^{\dagger}_ih_j \ket{\psi^0_{\boldsymbol{\nu'}}(N)}A_{\boldsymbol{\nu}'} \nonumber \\
\  E^{eff}a_{\boldsymbol{\mu}} &\simeq& s^2\sum_{\boldsymbol{\mu}'} \bra{\psi^0_{\boldsymbol{\mu}}(N-1)} \sum_{ij}h_ih^{\dagger}_j \ket{\psi^0_{\boldsymbol{\mu'}}(N-1)}   a_{\boldsymbol{\mu}'} \ ,
\end{eqnarray}
with
\begin{equation}\label{eq:eigenvalue-2Aviapprox}
E^{eff}(N,d)= [E-E^0_{low}(N,d) + E_F(N,d)] [E-E^0_{low}(N,d)]  \ .
\end{equation}
The eigenvalue equations (\ref{eq:eigenvalue-6Aiiiapprox}) are identical to the eigenvalue equations following from
\begin{eqnarray}\label{eq:eigenvalue-2Aivapprox}
\hat{H}^{eff}_N \ket{\phi_N} &=&  E^{eff}(N,d) \ket{\phi_N}  \nonumber \\
\hat{h}^{eff}_{N-1} \ket{\varphi_{N-1}}  &=& E^{eff}(N,d) \ket{\varphi_{N-1}}   \ .
\end{eqnarray}
with  the effective Hamiltonians
\begin{eqnarray}\label{eq:eigenvalue-2Avapprox}
\hat{H}^{eff}_N &=&  \tilde{s}^2  \frac{1}{d }\sum^{d}_{i,j=1}h^{\dagger}_i h_j \nonumber \\
\hat{h}^{eff}_{N-1} &=&  \tilde{s}^2  \frac{1}{d } \sum^{d}_{i,j=1}h_i h^{\dagger}_j \  \ ,
\end{eqnarray}
and $\ket{\phi_N}, \ket{\varphi_{N-1}}$  from Eq.~(\ref{eq:phi-varphi}).
Note, the eigenvalue $E^{eff}(N,d)$ is identical for $\hat{H}^{eff}_N $  and $\hat{h}^{eff}_{N-1}$.
$ \tilde{s}= (s/t) \sqrt{d}$  denotes the scaled  dimensionless coupling constant. We remind the reader
that we used $t=1$.
\\
\\
In the following it is more convenient to use the equivalence  $h^{\dagger}_i=S_i^{-}$, $h_i=S_i^{+}$,
$(1-2h^{\dagger}_ih_i)=2S_i^{z}$ between the hard-core Bose operators and the spin-one-half operators. The commutation relations of the latter read
\begin{equation}\label{eq:commutator}
\big[S_i^{+},S_j^{-}\big] = 2\delta_{ij} S_i^{z}, \quad
 \big[S_j^{z},S_i^{\pm}\big] = \pm \delta_{ij} S_i^{\pm} \ .
\end{equation}
The effective Hamiltonians become
\begin{eqnarray}\label{eq:eigenvalue-2Avapprox-spin}
\hat{H}^{eff}_N &=&  \tilde{s}^2  \frac{1}{d }\sum^{d}_{i,j=1} S^{-}_i S^{+}_j\\
\hat{h}^{eff}_{N-1} &=&  \tilde{s}^2  \frac{1}{d } \sum^{d}_{i,j=1}S^{+}_i S^{-}_j \  \ ,
\end{eqnarray}
Let $\vec{S}= \sum_{i=1}^d \vec{S}_i$ be the spin operator of the  total spin.
Because $\hat{H}^{eff}_N$(and $\hat{h}^{eff}_{N-1}$) commutes with $\vec{S}^2$ and $S^{z}$ all its eigenstates can be chosen such that they are also eigenstates of $\vec{S}^2$ and $S^{z}$ with eigenvalues $S(S+1)$ and $M$, respectively. They will be denoted by $\ket{S,M}$. $M$ is related to the particle number by $M=d/2-N$ and
for fixed $N$ the total spin quantum number takes the values $S=d/2-N,d/2-N+1,\cdots,d/2$. The corresponding eigenvalues of $\hat{H}^{eff}_N$ are given by
\begin{equation}\label{eq:E-eff(S,M)}
E^{eff}(S,M; \tilde{s}) = \tilde{s}^2 \frac{1}{d} \big[S(S+1)-M(M+1 \big]\ .
\end{equation}

The ground state eigenvalue $E_0(N,d;\tilde{s})$ of $\hat{H}$ follows from the largest eigenvalue $E^{eff}_{max}(N,d;\tilde{s})$ of $\hat{H}^{eff}_N$ which corresponds to $S_{max}=d/2$. Then we obtain from
Eq.~(\ref{eq:E-eff(S,M)}) in the thermodynamic limit $N \to \infty, d\to \infty$ with density $n=N/d$ fixed
\begin{equation}\label{eq:E-eff}
E^{eff}_{max}(N,d; \tilde{s}) \simeq  \tilde{s}^2N(1-n)\ .
\end{equation}
The corresponding eigenstates
are given by
\begin{eqnarray}\label{eq:states-eff}
\ket{\phi^{eff}_N} &=& \binom{d}{N}^{-1/2} (\sum^d_{i=1} h^{\dagger}_i)^N \ket{0} \nonumber \\
\ket{\varphi^{eff}_{N-1}} &=&  \binom{d}{N-1}^{-1/2} (\sum^d_{i=1} h^{\dagger}_i)^{N-1} \ket{0}  \ .
\end{eqnarray}
Using the spin analogy it is $\ket{\phi^{eff}_N} = \ket{d/2,d/2-N}$  and
$\ket{\varphi^{eff}_{N-1}} = \ket{d/2,d/2-(N-1)}$. Note, these eigenstates belong to the subspace with total momentum $Q=0$.

Substitution of $E^{eff}_{max}(N,d; \tilde{s})$ from Eq.~(\ref{eq:E-eff})  into Eq.~(\ref{eq:eigenvalue-2Aviapprox}) leads to the perturbed ground state eigenvalue
\begin{equation}\label{eq:groundapprox}
E_0(N,d;\tilde{s})  \simeq   E^0_{low}(N,d) - E_F(N,d)/2 - \sqrt{(E_F(N,d)/2)^2 +  \tilde{s}^2 d \ n(1-n)}  \  .
\end{equation}
We remind the reader that Eq.~(\ref{eq:E-Fermi})  implies  $E_F(N,d) \leq 0$ for $0 \leq n \leq 1/2$. Due to the particle hole duality $n \leq 1/2$ does not restrict generality.


The  energy, $E_1(N,d;\tilde{s})$, of the first excitation follows from Eq.~(\ref{eq:eigenvalue-2Aviapprox}) for the second largest effective eigenvalue.  Using the spin analogy  the latter corresponds to $S=S_{max}-1 \equiv d/2-1$ which yields $E^{eff}(N,d;\tilde{s})= E^{eff}_{max}(N,d;\tilde{s}) - \tilde{s}^2$. Accordingly we obtain
\begin{equation}\label{eq:first excitation}
E_1(N,d;\tilde{s}) \simeq   E^0_{low}(N,d) - E_F(N,d)/2 - \sqrt{(E_F(N,d)/2)^2 +  \tilde{s}^2 [d \ n(1-n) - 1 ]}  \  .
\end{equation}
It is easy to see that the excitation gap  $\Delta E(N,d;\tilde{s}) = E_1(N,d;\tilde{s}) - E_0(N,d;\tilde{s})$
is finite for $\tilde{s} \neq 0$.
The higher excitation energies $ E_n(N,d;\tilde{s})$ for $n=2, \cdots,N-1$ follow similarly using $S=d/2-n$. Therefore, the eigenvalues of $\hat{H}$ below $E^0_{low}(N,d)$ form a discrete spectrum of $N$
eigenvalues.
\\

Let us summarize:  The unperturbed spectrum in the subspace $Q=0$ consists of  two bands. One band, $B^0_{N-1}$, of  $K_{N-1}$ eigenvalues, $\{ E^0_{\boldsymbol{\mu}}(N-1,d) \}$, and the other band, $B^0_N$, with $K_N$ eigenvalues, $\{ E^0_{\boldsymbol{\nu}}(N,d) \}$). These two bands  correspond to $(N-1)$ and $N$ HCBs on the ring-lattice.  The band edges of $B^0_{N-1}$ are at  $\pm E^0_{low}(N-1,d)$ and those of $B^0_N$ at $\pm E^0_{low}(N,d)$. For $d$ finite, both sets $\{ E^0_{\boldsymbol{\mu}}(N-1,d) \}$ and $\{ E^0_{\boldsymbol{\nu}}(N,d) \}$ are disjoint, and for density $n=N/d \leq 1/2$   $B^0_{N-1}$ is  a subset within the interval $[-|E^0_{low}(N,d)|,+|E^0_{low}(N,d)|]$.
Turning on $s$ leads to a coupling between these two bands. Part of these two bands persist. The lower  band edge of the perturbed band coincides for $d= \infty$ with   the lower band edge of the unperturbed band.
Below that band a discrete spectrum of  maximally $N$ eigenvalues  occurs exhibiting an excitation gap. The number of discrete eigenvalues may change with $\tilde{s}$.

The number $N_0= (1/d)\bra{\Psi_N} \sum^d_{i,j=1} h^{\dagger}_i h_j \ket{\Psi_N}$ of condensed particles in the state $ \ket{\Psi_N}$ is easily obtained since
$(1/d)\bra{\Psi_N} \sum^d_{i,j=1} h^{\dagger}_i h_j \ket{\Psi_N} \simeq \bra{\Psi_N} \hat{H}^{eff}_N \ket{\Psi_N}/\tilde{s}^2 $. Substitution of $\ket{\Psi_N}$ from Eq.~(\ref{eq:Psi-Suppl})  leads to
$N_0= [ |\alpha|^2 \bra{\phi^{eff}_N} \hat{H}^{eff}_N \ket{\phi^{eff}_N} +  |\beta|^2 \bra{\varphi^{eff}_{N-1}} \hat{H}^{eff}_{N} \ket{\varphi^{eff}_{N-1}}]/\tilde{s}^2$. Making use of the commutation relations of $\{h^{\dagger}_i\}, \{h_j\}$ one can express $\hat{H}^{eff}_{N} $ in the second summand  by $\hat{h}^{eff}_{N-1}$:
\begin{equation}\label{eq:condensate-Ai}
   \hat{H}^{eff}_N=\hat{h}^{eff}_{N-1} - \frac{\tilde{s}^2}{d}\sum^d_{i=1}(1-2\hat{n}_i)   \  .
\end{equation}
With $|\alpha|^2 +  |\beta|^2 =1$, \ $ \sum^d_{i=1}\hat{n}_i   \ket{\varphi^{eff}_{N-1}} = (N-1) \ket{\varphi^{eff}_{N-1}}$,  Eqs.~(\ref{eq:eigenvalue-2Aivapprox}),
(\ref{eq:eigenvalue-2Avapprox}) and   (\ref{eq:E-eff}) we obtain for  $d \to \infty$ the final result
\begin{equation}\label{eq:condensate-Aii}
N_0(N;\tilde{s}) \simeq N \big[(1-n) -  |\beta|^2(1-2 n) N^{-1} \big]  \  .
\end{equation}
Since $ |\beta|^2 \leq 1$ the second term on the r.h.s. of Eq.~(\ref{eq:condensate-Aii}) is a negative
correction(for $n < 1/2$)  to the leading order term $(1-n)$ which is of order $\mathcal{O}(1/N)$.
\\

As discussed above the mapping of the original model to an effective one is valid if one is allowed to replace $ E^0_{\boldsymbol{\nu'}}(N,d)$ by the unperturbed ground state energy eigenvalue
$E^0_{low}(N,d)$. This is equivalent to the replacement of $\hat{H}_0$ in the first line of Eq.~(\ref{eq:eigenvalue-1A}) by $E^0_{low}(N,d)$. With this replacement and that of $(\ket{\phi_N}, \ket{\varphi_{N-1}})$
by $(\ket{\phi^{eff}_N}, \ket{\varphi^{eff}_{N-1}})$ from Eq.~(\ref{eq:states-eff}) one can solve the linear equation for $\alpha, \beta$.  With use of $ [E_0(N,d;\tilde{s})-E^0_{low}(N,d)]$ from Eq.~(\ref{eq:groundapprox}) and the normalization condition $|\alpha |^2+|\beta|^2=1$ one obtains for the ground state
\begin{eqnarray}
|\alpha|^2 &\simeq & 1 -\frac{1}{2}\tilde{s}^2N(1-n) \Big\{ \big[(E_F/2)^2+\tilde{s}^2N(1-n) \big] - (E_F/2) \sqrt{(E_F/2)^2+\tilde{s}^2N(1-n)} \Big\}^{-1} \label{eq:alpha-ground}  \\
|\beta|^2 &\simeq& \frac{1}{2}\tilde{s}^2N(1-n)\Big\{ \big[(E_F/2)^2+\tilde{s}^2N(1-n) \big] - (E_F/2) \sqrt{(E_F/2)^2+\tilde{s}^2N(1-n)} \Big\}^{-1}. \label{eq:beta-ground}
\end{eqnarray}
%
%
%

Now we discuss the validity of the above  mapping of the original eigenvalue problem to an effective one.
The simplest   limiting case under which the mapping  becomes exact  is the strong coupling limit $s \to \infty$.
For the ground state energy $E=E_0(N,d;\tilde{s})$ the denominators in Eq.~(\ref{eq:eigenvalue-3A}) can be rewritten as follows
\begin{equation}\label{eq:denom}
 [E_0(N,d;\tilde{s})-E^0_{\boldsymbol{\nu'}}(N,d)] = [E_0(N,d;\tilde{s})-E^0_{low}(N,d)] \{1 + [E^0_{\boldsymbol{\nu'}}(N,d) - E^0_{low}(N,d)]/ [E^0_{low}(N,d) - E_0(N,d;\tilde{s})]\}^{-1} \  .
\end{equation}
and similar for $[E_0(N,d;\tilde{s}) -E^0_{\boldsymbol{\mu'}}(N-1,d)]$. Accordingly, the mapping becomes exact if

 \begin{equation}\label{eq:approx-validity}
  [E^0_{\boldsymbol{\nu'}}(N,d) - E^0_{low}(N,d)]/ [E^0_{low}(N,d) - E_0(N,d;\tilde{s})] \to 0\  .
\end{equation}
Since the upper band edge fulfils $E^0_{up}(N,d) = - E^0_{low}(N,d)$ it follows $|[E^0_{\boldsymbol{\nu'}}(N,d) - E^0_{low}(N,d)]\ \leq 2 \ |E^0_{low}(N,d)]|$ for all $\boldsymbol{\nu'}$ and the   condition (\ref{eq:approx-validity}) becomes
$2  |E^0_{low}(N,d)]|/ [E^0_{low}(N,d) - E_0(N,d;\tilde{s})] \ll 1$. Using Eqs.(\ref{eq:E-0-a}),
 (\ref{eq:E-Fermi}) and (\ref{eq:groundapprox}) we obtain the condition


\begin{equation}\label{eq:strong-coupling}
s \gg \frac{4}{\pi} \frac{\sin{\pi}n}{\sqrt{n(1-n})}  \  .
\end{equation}

The reason why the  mapping becomes exact in the scaling limit   $s \to 0, d \to \infty$ with $\tilde{s}=s \sqrt{d}$ and $N$ fixed, is more subtle.  In that case the variation  with $\boldsymbol{\mu'}$ of  the numerator  $b^{\dagger}_{\boldsymbol{\nu}\boldsymbol{\mu'}}b_{\boldsymbol{\mu'}\boldsymbol{\nu'}}$  and of the denominator $[E-E^0_{\boldsymbol{\mu'}}(N-1,d)]$ in the first line of Eq.~(\ref{eq:eigenvalue-3A})  plays the essential role.  In Appendix B we prove that $b_{\boldsymbol{\mu}\boldsymbol{\nu}} \sim \sqrt{d}$ for $\boldsymbol{\mu}, \boldsymbol{\nu}$ fixed and  $d \to \infty$ whereas $b_{\boldsymbol{\mu}\boldsymbol{\nu}} = \mathcal{O}(1)$ if $\mu_k \in \boldsymbol{\mu}$ and  $\nu_k \in \boldsymbol{\nu}$
are of $\mathcal{O}(d)$. This means that $b_{\boldsymbol{\mu}\boldsymbol{\nu}}$
decreases fast with increasing  $\mu_k$ and  $\nu_k$. Therefore the main contributions  in the sums in
Eq.~(\ref{eq:eigenvalue-3A})  for $N$ arbitrary large but fixed comes from $\boldsymbol{\mu'},\boldsymbol{\nu}$  with $\mu'_k$ and  $\nu_k$ arbitrary large but fixed. Therefore restricting
the sums in Eq.~(\ref{eq:eigenvalue-3A}), e.g., over $\boldsymbol{\mu'}$  to $|\mu'_k| \leq \sqrt{d}$ for all $k$  does not change the result if $d$  becomes very large. Due to this restriction of $\{ \mu_k \}$ we obtain with
$E^0_{low}(N,d) \simeq -2N$ and $\cos{(2\pi \mu_k/d)} \leq 1-  2\pi^2/d$ from
 (\ref{eq:E-0}) the upper bound $ [E^0_{\boldsymbol{\nu'}}(N,d) - E^0_{low}(N,d)] \ll 4 \pi^2 N/d$ for the numerator in   Eq.~(\ref{eq:approx-validity}). Substituting this upper bound and   $[E^0_{low}(N,d) - E_0(N,d;\tilde{s})]$ from Eq.~(\ref{eq:groundapprox}) into Eq.~(\ref{eq:approx-validity}) leads for $N$ fixed and $d \gg1$   to the condition
\begin{equation}\label{eq:scaling limit}
s \gg    2\sqrt{2} \pi /d\  .
\end{equation}
Here we also used  $E_F(N,d) \simeq -2$  because $n \approx 0$   for $N$ fixed and $d \gg 1 $(cf. Eq.~(\ref{eq:E-Fermi})).

\subsection*{Behavior of $b_{\mu,\nu}$  for $N$ fixed and $d \to \infty$}
To study the behavior of  $b_{\boldsymbol{\mu}\boldsymbol{\nu}}$ for $N$ fixed and  $d \to \infty$ we first observe that the translational invariance on the ring implies that Eq.~(\ref{eq:eigenvalue-4A}) becomes $b_{\boldsymbol{\mu}\boldsymbol{\nu}} = d \  \bra{\psi^0_{\boldsymbol{\mu}}(N-1)} h_1 \ \ket{\psi^0_{\boldsymbol{\nu}}(N)}$.
Substituting $\ket{\psi^0_{\boldsymbol{\mu}}(N-1)}$, $\ket{\psi^0_{\boldsymbol{\nu}}(N)}$  from Eq.~(\ref{eq:psi-0-0}) and taking advantage of the ordering $1 \leq n_1< \cdots < n_N \leq d$ one arrives at
\begin{equation}\label{eq:eigenvalue-12Ai}
 b_{\boldsymbol{\mu}\boldsymbol{\nu}} = d  \sum_{2 \leq m_2 < \cdots < m_N \leq d} \psi^0_{\boldsymbol{\mu}}(m_2,\cdots,m_N)^{*} \psi^0_{\boldsymbol{\nu}}(1,m_2,\cdots,m_N)    \  .
\end{equation}
Introducing new variables $n_i=m_{i+1}-1$ and taking the translational invariance into account this yields
\begin{equation}\label{eq:eigenvalue-12Aii}
 b_{\boldsymbol{\mu}\boldsymbol{\nu}} = d  \sum_{1 \leq n_1 < \cdots < n_{N-1} \leq d-1} \psi^0_{\boldsymbol{\mu}}(n_1,\cdots,n_{N-1})^{*}  \psi^0_{\boldsymbol{\nu}}(0,n_1,\cdots,n_{N-1})    \  .
\end{equation}
Substituting the  normalized `wave functions' from Eq.~(\ref{eq:psi-0}) leads to
\begin{equation}\label{eq:eigenvalue-6Aiii}
b_{\boldsymbol{\mu}\boldsymbol{\nu}} = d d^{-(N-\frac{1}{2})}  \sum_{P \in S_{N-1}}  \sum_{P' \in S_N} sgn(P) sgn(P')
\sum_{1 \leq n_1 < \cdots < n_{N-1} \leq (d-1)} \exp{\Big[-i \sum^{N-1}_{k=1} \big(q_{\mu_{P(k)}} - q_{\nu_{P'(k+1)}}   \big)n_k\Big]} \  .
\end{equation}
The crucial quantity is the 2nd line of Eq.~(\ref{eq:eigenvalue-6Aiii}). This sum can be written as
$ \sum^{d-N+1}_{n_1=1} \sum^{d-N+2}_{n_2=n_1+1} \cdots  \sum^{d-1}_{n_{N-1}=n_{N-2}+1} (\cdots)$.
Each single sum generates a denominator of the form
$\big(1- \exp[-i(\sum^{N-1}_{j=1} a_jq_{\mu_j} -\sum^{N}_{j'=1} b_{j'}q_{\nu_{j'}})]\big)$
where the integers $\{a_j\}$ and $\{ b_{j'}\}$ take values $0, \pm 1$. There is a product of $(N-1)$ such denominators. For $\boldsymbol{\mu},\boldsymbol{\nu}$ fixed and for $d \to \infty$ this product is proportional to $d^{N-1}$
. Performing the sums in the 2nd line of Eq.~(\ref{eq:eigenvalue-6Aiii}) also generates numerators
of the form $\big(1-\exp[-i(\sum^{N-1}_{j=1} a'_jq_{\mu_j} -\sum^{N}_{j'=1} b'_{j'}q_{\nu_{j'}} )d] \big)$ where
$\{a'_j\}$ and $\{ b'_j\}$ take values $0, \pm 1$. $\exp[iq_{\mu_j}d]=+1(-1)$ and $\exp[iq_{\nu_j}d]=-1(+1)$ for $N$ even(odd). Therefore some of the  numerators vanish and some do not. The latter
take the value $2$.  Accordingly, for $d \to \infty$ the contribution of these terms in the 2nd line of Eq.~(\ref{eq:eigenvalue-6Aiii}) is of order $d^{N-1}$. The contribution of all the other terms are  of $\mathcal{O}(d^{N-2})$. Taking the prefactor  $d d^{-(N-\frac{1}{2})}$ on the r.h.s. of Eq.~(\ref{eq:eigenvalue-6Aiii}) into account one obtains for  $(\boldsymbol{\mu},\boldsymbol{\nu})$ arbitrary but fixed and $d \to \infty$ in leading order in $d$
\begin{equation}\label{eq:eigenvalue-12Aiv}
 b_{\boldsymbol{\mu}\boldsymbol{\nu}} \sim  \sqrt{d}      \ ,
\end{equation}
which we wanted to prove.

\end{document}